\documentclass[twocolumn]{aastex63}
\usepackage{caption}
\usepackage[skip=0.5ex]{subcaption}
\usepackage{amsmath,amstext}
\usepackage[T1]{fontenc}
\usepackage{apjfonts}
\usepackage[figure,figure*]{hypcap}
\usepackage{multirow}
\usepackage{subfiles}
\usepackage{graphicx}
\DeclareGraphicsExtensions{.pdf,.png}

\defcitealias{Griffith23}{G23}
\defcitealias{Kirby2018_data}{K18}

\begin{document}

\title{Measuring Dwarf Galaxy Intrinsic Abundance Scatter with Mid-resolution Spectroscopic Surveys:\\Calibrating APOGEE Abundance Errors}
\correspondingauthor{Jennifer Mead}
\email{jennifer.mead@columbia.edu}


\author[0009-0006-4744-2350]{Jennifer Mead}
\affiliation{Department of Astronomy, Columbia University, New York, NY 10027, USA}

\author[0000-0001-5082-6693]{Melissa Ness} 
\affiliation{Research School of Astronomy \& Astrophysics, Australian National University, Canberra ACT 2611, Australia}
\affiliation{Department of Astronomy, Columbia University, New York, NY 10027, USA}

\author[0000-0003-3479-4606]{Eric Andersson}
\affiliation{Department of Astrophysics, American Museum of Natural History, New York, NY 10024, USA}

\author[0000-0001-9345-9977]{Emily J. Griffith}
\affiliation{Center for Astrophysics and Space Astronomy, Department of Astrophysical and Planetary Sciences, University of Colorado, 389 UCB, Boulder, CO 80309-0389,
USA}
\altaffiliation{NSF Astronomy and Astrophysics Postdoctoral Fellow}

\author[0000-0003-1856-2151]{Danny Horta}
\affiliation{Center for Computational Astrophysics, Flatiron Institute, 162 5th Avenue, New York, NY 10010, USA}


\begin{abstract}
The first generations of stars left their chemical fingerprints on metal-poor stars in the Milky Way and its surrounding dwarf galaxies.  While instantaneous and homogeneous enrichment implies that groups of co-natal stars should have the same element abundances, small amplitudes of abundance scatter are seen at fixed [Fe/H].  Measurements of intrinsic abundance scatter have been made with small, high-resolution spectroscopic datasets where measurement uncertainty is small compared to this scatter. In this work, we present a method to use mid-resolution survey data, which has larger errors, to make this measurement.  Using APOGEE DR17, we calculate the intrinsic scatter of Al, O, Mg, Si, Ti, Ni, and Mn relative to Fe for 333 metal-poor stars across 6 classical dwarf galaxies around the Milky Way, and 1604 stars across 19 globular clusters.  We calibrate the reported abundance errors in bins of signal-to-noise and [Fe/H] using a high-fidelity halo dataset.  Applying these calibrated errors to the APOGEE data, we find small amplitudes of average intrinsic abundance scatter in dwarf galaxies ranging from 0.03~$-$~0.09~dex with a median value of 0.047~dex.  For the globular clusters, we find intrinsic scatters ranging from 0.01~$-$~0.11~dex, with particularly high scatter for Al and O. Our measurements of intrinsic abundance scatter place important upper bounds, which are limited by our calibration, on the intrinsic scatter in these systems, as well as constraints on their underlying star formation history and mixing, that we can look to simulations to interpret.
\end{abstract}

\section{Introduction}
Quenched dwarf galaxies that are bound to and inhabit our stellar halo serve as laboratories for studying snapshots of the early chemical evolution of the universe. With average stellar metallicities typically $\rm{[Fe/H]} < -1$ \citep{Kirby2008, Kirby2013, Geha2013,Li2018,Fu2023}, their oldest, most metal-poor stars contain the chemical fingerprints of the very first generations of stars, including Population III \citep{Frebel2012, Ji2015, Hartwig2018, Hartwig2019}.  An understanding of the contribution to chemical enrichment from Population II stars is also necessary for placing constraints on the chemical evolution of the early universe.  To decode the signatures of the chemical ancestry of dwarf galaxies, it is essential to understand the observational effects of the propagation of elements from various populations of stars and from a variety of nucleosynthetic sources throughout the interstellar medium \citep[ISM;][]{Ji2015}.

The chemical history of a stellar population is observed through measurements of element abundances, as well as their intrinsic abundance scatter -- the true scatter of the data, accounting for measurement uncertainties -- which necessitates careful consideration of uncertainties on abundances.  Previous studies have uncovered overall element abundance scatter in numerous abundances, both within and between dwarf galaxies \citep{Hill2019, Ji2020, Ji2022}, but there remains a need to resolve the intrinsic scatters of chemical abundances above the level of observational errors; a difficult task for even small high-resolution spectroscopic studies.  Mid-resolution spectroscopic surveys [e.g. APOGEE \citep{apogee_overview}, GALAH \citep{GALAH}, Gaia RVS \citep{GaiaMission,GaiaDR3} and LAMOST \citep{LAMOST}] afford the community with copious amounts of data, and target numerous stellar systems. Nonetheless, measurements of the intrinsic abundance scatter are made even more challenging by medium-resolution data having larger measurement uncertainties than data from higher signal-to-noise (S/N), higher resolution data. Additionally, medium-resolution data may have potentially inaccurate estimates of measurement uncertainty in some parameter spaces. An important question to ask, however, is if we can use large mid-resolution surveys, to test the the intrinsic scatter of metal-poor systems that have been targeted and furthermore if we can validate or if necessary, calibrate, the uncertainty estimates so that these are accurate.

From a theoretical perspective, the intrinsic scatter of element abundances in any system of stars is dependent on a number of properties. This includes the efficiency of gas mixing and metal diffusion in the ISM \citep{Krumholz2018, Escala2018, Emerick2020, Ji2023}, metal-retention and recycling through inflows and outflows \citep[][Mead et al. in prep]{Emerick2018}, stochastic star formation \citep{Welsh2021, Pan2023, Griffith23}, mass and metallicity dependence of nucleosynthetic sources \citep{Muley2021}, and the assembly history of the system \citep{Genina2019,Patel2022}.  The assessment of the scatter of individual elements is one indirect approach to access these environmental parameters.

The interpretation of the intrinsic scatter for an ensemble of elements is a powerful link to the details of the formation and enrichment histories of dwarf galaxies, star clusters, and stellar halos.  For example, discoveries in the last few decades have found multiple populations of stars in globular clusters, and thus larger intrinsic scatter in some abundance dimensions \citep[][and references therein]{Bastian2018}, hinting at a more complex formation pathway than previously thought.  Similarly, in dwarf galaxies the magnitude of the intrinsic scatter provides clues to the mixing of chemical yields and stochasticity of star formation, particularly in the early Universe.  In the halo, which is thought to be composed of globular clusters and disrupted dwarfs, the intrinsic scatter may provide information on where the stellar populations of the halo originated from.

Because different nucleosynthetic channels contribute to the enrichment of different elements, intrinsic scatter is also a metric by which we can study the contribution of different nucleosynthetic sources to the chemical evolution of different stellar systems, as well as the mass and metallicity dependence of yields.  For example, light elements such as C, N, and F are primarily produced by low and intermediate mass stars on the asymptotic giant branch (AGB) \citep{Karakas2010,Kobayashi2011b}.  Notably, some element abundances such as C, N, and O are sensitive to the evolutionary state of the star, in particular, mixing that brings the products of the CNO-cycle to the photosphere during the RGB phase.  The $\alpha$-elements (e.g. O\footnote{O is also produced through the CNO cycle, and so not always considered an $\alpha$-element, but this contribution is subdominant for low-metallicity stars.}, Mg, Si, Ca, Ti) are primarily produced by massive stars and ejected by core-collapse (Type II) supernovae \citep[SNe;][]{Timmes1995,Kobayashi2006}.  About half of Fe-peak elements (e.g. Cr, Mn, Fe, Ni, Co) are produced in Type Ia supernovae \citep{Kobayashi2009}.  Putting these together, the well-known [$\alpha$/Fe]-[Fe/H] relationship, which shows the delay in enrichment in Fe from Type Ia SNe versus the short-timescale production of $\alpha$-elements by Type II SNe, can be used to characterize the star formation rates in galaxies \citep[e.g.][]{Tinsley1980,Matteucci1986,Hayden2015}.

Other elements, such as the odd-Z elements (e.g. Al, P) and neutron-capture elements (s- and r-process elements) depend on the excess of neutrons available.  Odd-Z elements depend on the metallicity of the progenitor star as their formation requires a surplus of neutrons made from $^{22}$Ne during He-burning \citep{Kobayashi2011b}.  Strong s-process elements (Sr to Pb) are produced in low-mass AGB stars \citep{Busso1999,Herwig2005,Karakas2014}, whereas weak s-process elements (between Fe and Sr) are produced in massive stars near solar metallicity \citep{Pignatari2010}.  The formation mechanism for r-process elements is debated but neutron star mergers are a likely scenario \citep{Lattimer1974,Rosswog1999,Goriely2011,Wanajo2014}.

\citet{Kobayashi2020} developed galactic chemical evolution models to place constraints on the origin of elements as a function of time and galactic environment.  Models such as this, along with an understanding of gas mixing, the ratio of nucleosynthetic sources, and stochasticity of star formation can be used to predict the origin and amount of scatter in dwarf galaxy observations. High-resolution star-by-star models with individual stellar feedback \citep[e.g. Brauer et al. in prep;][]{Andersson+2023} are likewise uniquely poised to explore these physical drivers in the context of intrinsic abundance scatter.

Measurements of the intrinsic scatter have been successfully made for different systems and structures. Intrinsic abundance scatters have been calculated in both the globular cluster systems of the stellar halo, and open cluster systems in the disk. Metal-rich open star clusters, which are thought to be the building blocks of the disk, and born together from the same molecular cloud, are near chemically homogeneous. Studies measure small amplitudes of open cluster element abundance intrinsic scatter, on the order of $<$ 0.03 dex \citep[e.g.][]{Bovy2016, Ness2018,Poovelil2020}.  In the disk itself, scatter in [O/Fe] has been shown to be 0.03-0.04 dex in the high- and low-$\alpha$ disk \citep{Bertran_de_Lis2016}. Several studies also identify non-zero intrinsic abundance scatters in globular clusters, in particular, this is most significant for light elements (for example, He has a scatter of up to 0.1 dex) \citep{Milone2018,Meszaros2020}, but is also seen to a lesser extent in heavy elements \citep{Gratton2020}.

In the solar neighborhood, \citet[hereafter \citetalias{Griffith23}]{Griffith23} measured element abundances and intrinsic abundance scatters for metal-poor stars for 12 elements ranging from 0.04 for [Cr/Fe] to 0.16 for [Na/Fe] dex.  Because of their low metallicity, these stars are likely field halo stars.  They notably show that these scatters can be produced by N$\sim$50 core-collapse supernovae, which is a number too small to accurately represent an initial mass function (IMF).

\begin{figure*}[t!]
    \begin{minipage}[t!]{.99\textwidth}
    \centering
    \includegraphics[width=\linewidth]{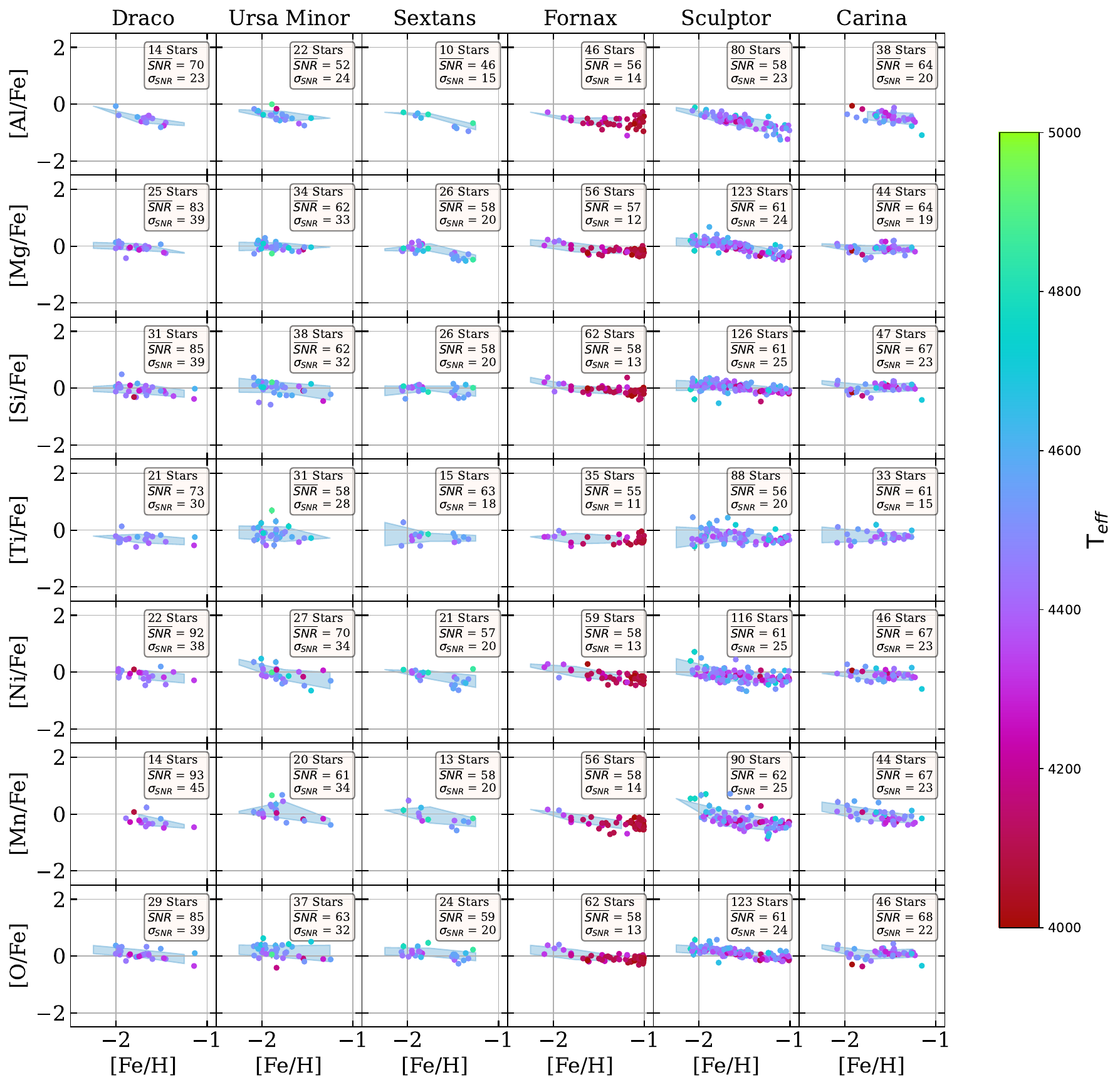}
    \caption{Measurements of element abundances relative to Fe for stars in our sample with S/N $\geq$ 25.  Points are colored by $\rm{T_{eff}}$. The blue shaded region represents 1$\sigma$ around the mean in 3 equally spaced bins in [Fe/H] from -2.1 to -1.}
    \label{fig:data_summary}
    \end{minipage}
\end{figure*}

Quantifying the scatter in co-natal (similar [Fe/H]) stellar populations serves as a baseline for understanding the temporal, chemical, and mechanical assembly of these systems.  In the case of dwarf galaxies, this accesses the earliest modes of star formation, as well as the earliest epochs of Milky Way assembly. For stellar clusters, this quantification has implications for the feasibility of chemical tagging \citep{Freeman2002}, or the use of chemical abundance patterns to identify common birth sites of stars.

\begin{figure*}[t!]
    \begin{minipage}[t!]{.99\textwidth}
    \centering
    \includegraphics[width=\linewidth]{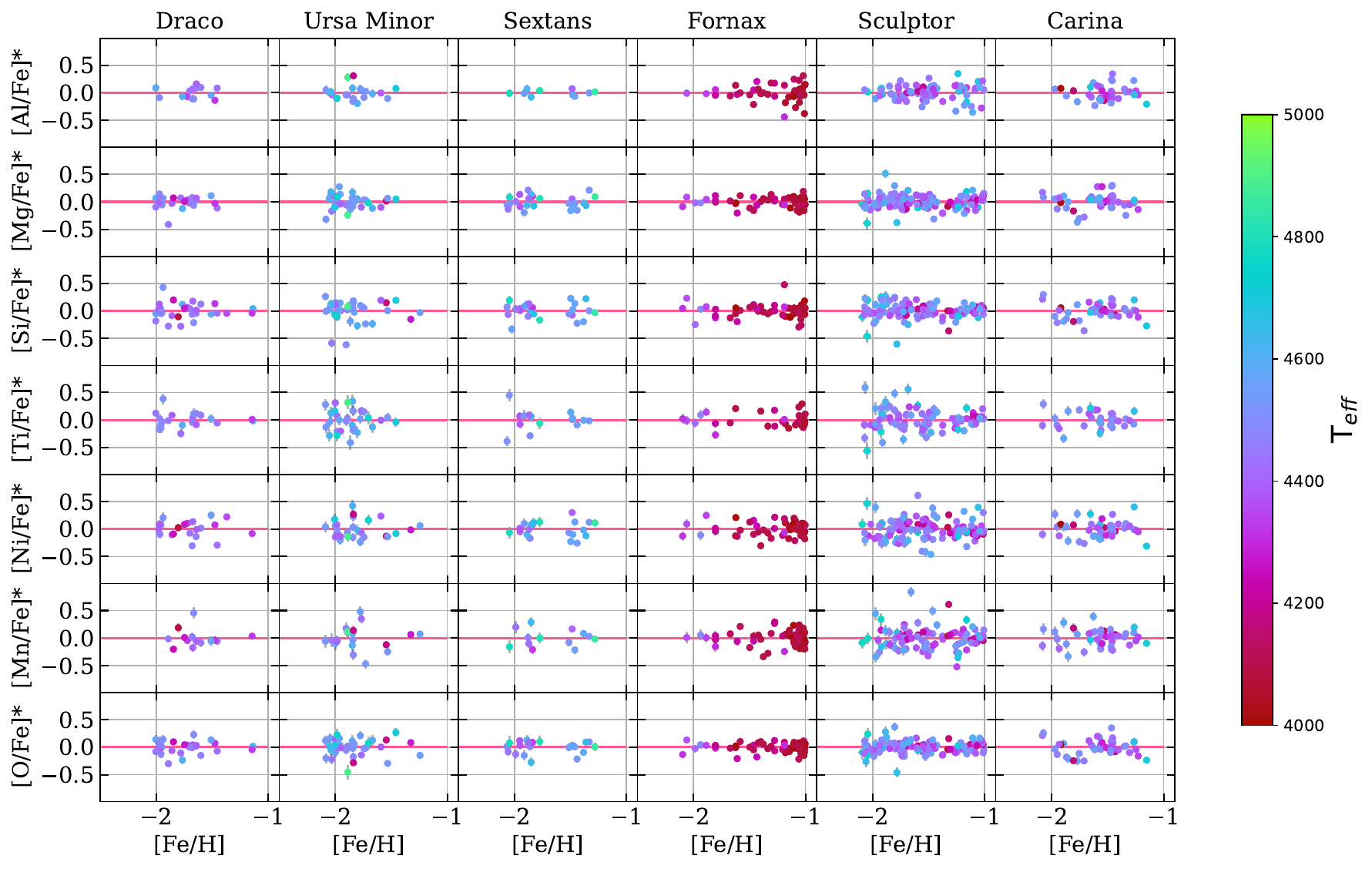}
    \caption{Same as Figure \ref{fig:data_summary} with best fit line subtracted to remove scatter dependence on [Fe/H].  $\rm{[X/Fe]^*} = \rm{[X/Fe]} - \rm{\langle{[X/Fe]}\rangle}$. }
    \label{fig:data_res}
    \end{minipage}
\end{figure*}

In our study, we use the SDSS \citep{SDSS} Apache Point Observatory Galactic Evolution Experiment (APOGEE) survey \citep{apogee_overview}  DR17 \citep{SDSSDR17} to study the intrinsic abundance scatter of metal-poor ([Fe/H] $\le$ -1) stars in dwarf galaxies surrounding the Milky Way.  Previous works have studied the abundances of the Milky Way satellites with APOGEE data \citep[e.g.][]{Hasselquist2021,Fernandes2023}.  The APOGEE survey has IR spectra (R=22,500) for nearly 734,000 stars in and around the Milky Way, including 7 targeted dwarf spheroidal galaxies (dSph): Draco, Ursa Minor, Bootes I, Sextans, Fornax, Sculptor, and Carina (total 718 stars).  The spectra were fit for stellar parameters and chemical abundances using the APOGEE Stellar Parameter and Chemical Abundances Pipeline (ASPCAP) \citep{aspcap}. In this work, we focus on elements from the $\alpha$, Fe-peak, and light nucleosynthetic families.

The layout of this paper is as follows: in Section \ref{sec:data} we discuss the sample of APOGEE data that we use, along with checks applied to the errors and the spectra themselves to estimate the uncertainty of their quality.  In Section \ref{sec:calib}, we present a method for calibrating the uncertainties on the APOGEE abundance data using intrinsic abundance scatters derived from high fidelity data.  Section \ref{sec:scatter} contains measurements and results of intrinsic abundance scatter using the calibrated APOGEE abundance uncertainties.  Lastly, we conclude in Section \ref{sec:conc} and discuss the implications of our results.

\section{Data} \label{sec:data}
In our work, we use APOGEE DR17 to study metal-poor red giant stars in three types of objects: dSphs, globular clusters (GC), and the Milky Way stellar halo.

\subsection{Data Cuts}
For all objects, we implement a metallicity cut of $-2.1 \le$ [Fe/H] $\le -1$ to obtain the metal-poor stars whose scatter we are interested in. For all but the halo stars, we identify likely members of each system using the assigned APOGEE \texttt{MEMBERFLAG}.  Halo stars are identified as metal-poor stars with distance $d \le 2500 \rm{pc}$.  As in \citetalias{Griffith23}, these stars are from the solar neighborhood, but due to their low-metallicity, are likely halo stars.  We remove stars in known substructure from this sample including Gaia-Sausage-Enceladus (GSE) using the same mask as \citet{Horta2023}, and globular clusters using the value-added catalog from \citet{GC_VAC_APOGEE_DR17}.  We also include a stricter cut on GSE stars by eliminating high eccentricity (\textit{e} $>$ 0.7) stars, but this had no effect on our results compared to a sample that omits this cut. 

To ensure the quality of the data, we remove all stars which were flagged with warnings generated by ASPCAP for their stellar parameters (\texttt{STAR\_BAD}) or abundance measurements (\texttt{M\_H\_BAD}; \texttt{ELEMFLAG}).  To exclude very low-quality spectra, we only use stars with a $\rm{S/N} \ge 25$, with the final sample for dSphs having a median $\rm{S/N} = 61.22$.  Lastly, we only include stars with $4000 \le \rm{T_{eff}} \le 5000 K$ (median $\rm{T_{eff}} = 4434 \, \rm{K}$ for dSphs) and $0.0 \le \rm{log}g \le 3.0$ (median $\rm{log}g = 0.99$ for dSphs), keeping the range wide enough to contain as much of the data as possible but narrow enough to avoid systematic variations of these quantities with [X/Fe] \citep[e.g.][]{Griffith2021a,Weinberg2022}.  Our final sample includes 333 stars in dSphs, 1604 stars in GCs and 291 stars in the halo.  The Appendix includes a summary of the resulting number of stars in our sample for each object we study.  Because Bootes I has so few metal-poor stars at the required S/N, we exclude Bootes I from analysis.

\subsection{Data Summary}
We use chemical abundances derived by ASPCAP for the stars in our sample \citep{aspcap}.  We focus on the following elements: Al (light elements); O, Mg, Si, Ti ($\alpha$-elements); and Ni and Mn (Fe-peak elements).  Figure \ref{fig:data_summary} shows the abundance [X/Fe] and metallicity [Fe/H] measurements for stars in our dSph sample.  Data points in Figure \ref{fig:data_summary} are colored by $\rm{T_{eff}}$. Generally, we see negative trends in [X/Fe] with [Fe/H]. As we are interested in processes that influence scatter in abundances for stars at the same metallicity, or in other words, the scatter around the trend line, we fit and subtract a second order model to the data in the [Fe/H]-[X/Fe] plane.  We also account for trends in $\rm{T_{eff}}$ that may be expected for elements such as O, which are sensitive to the evolutionary state of the star, by including a linear term in our model as we want the measured scatter to be independent of stellar parameters. We perform a multivariate linear regression over the full model (Equation \ref{eq:best_fit_model}) to fit coefficients $a-e$.
\begin{equation} \label{eq:best_fit_model}
    \rm{\langle[X/Fe]\rangle} = \rm{a[Fe/H]} + \rm{b[Fe/H]}^2 + \rm{cT_{eff}} + \rm{d([Fe/H]\times T_{eff})} + \rm{e}
\end{equation}
We find that the absolute gradients of T$\rm{_{eff}}$ for this regression are on the order $10^{-3}$ dex K$^{-1}$, ranging from $8 \times 10^{-4}$ to $3.5 \times 10^{-3}$.  Although the gradients appear small, over a 500K range, which is typical for a given [X/H]-dSph pair in our sample, they suggest a large variation of $\sim$0.5 dex.  However, this apparent variation may not represent the true gradient due to the influence of the cross term.  To demonstrate that the effect of T$\rm{_{eff}}$ is in fact small, we report that the average $\Delta$[X/Fe] over the range of T$\rm{_{eff}}$ for each [X/Fe]-dSph pair is only 0.18 dex for a model that does not include T$\rm{_{eff}}$ terms.

Returning to the model in Equation \ref{eq:best_fit_model}, the gradients for the cross term are a factor of $\sim 2$ less than those for T$\rm{_{eff}}$.  We nonetheless include the term as it appears to remove correlations between [Fe/H] and T$\rm{_{eff}}$, which may be an artifact of the data processing steps taken to produce the abundance catalog or the lack of data in the low-metallicity regime. In Figure \ref{fig:data_res}, we show the resulting residuals for the dSphs, revealing the scatter of our sample with [Fe/H].  We similarly do this for the Milky Way stellar halo and GC stars in our sample.  This trend-subtracted data is used in all further analysis.

\begin{figure*}
    \centering
    \includegraphics[width=0.9\linewidth,trim={1cm 3.5cm 1cm 4cm},clip]{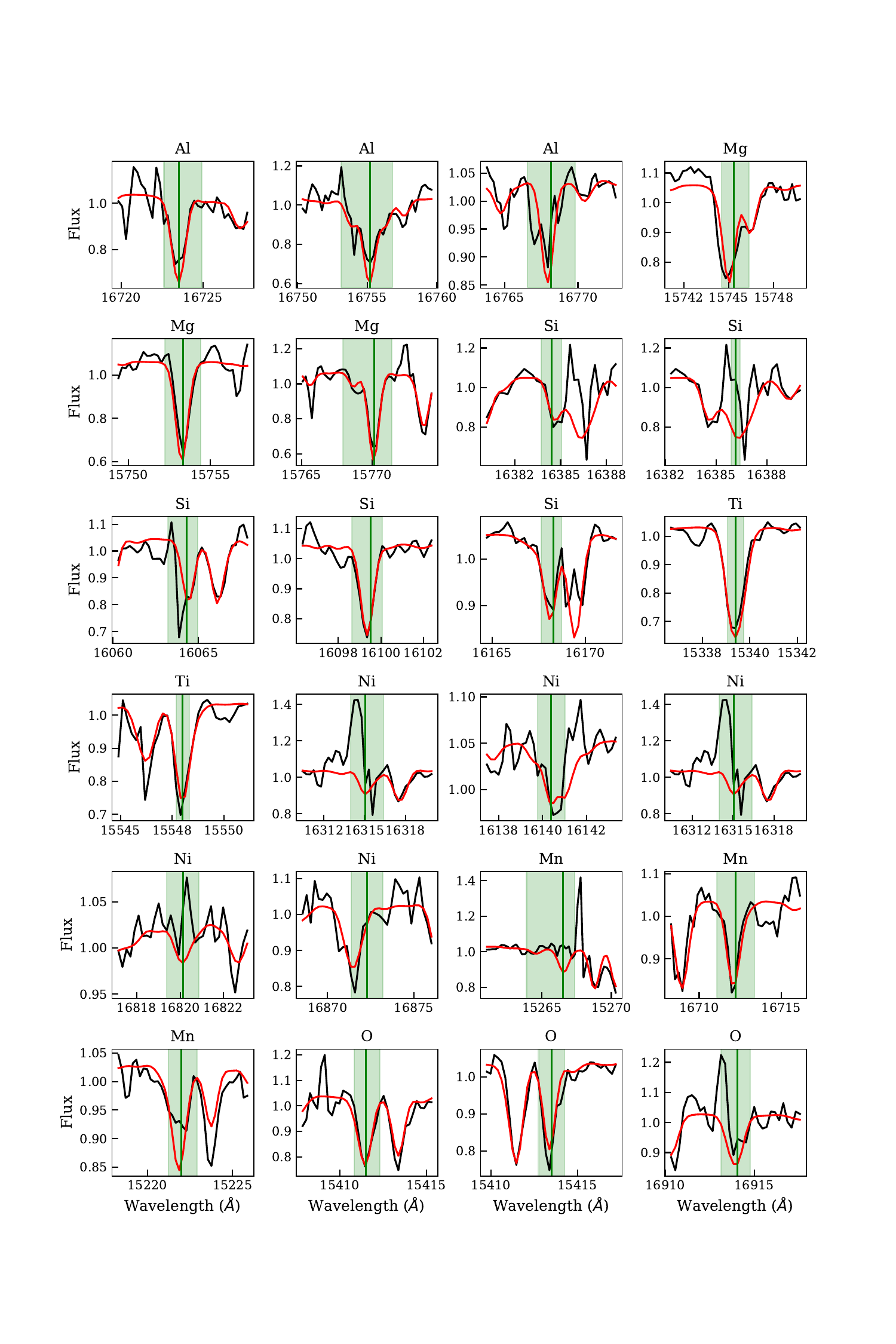}
    \caption{Example line windows for star 2M02383095-3456544 in Fornax, which has an S/N of 58.6.  The combined spectrum is shown in black with the ASPCAP model in red.  The line window is indicated by the green shading.}
    \label{fig:spec_ex}
\end{figure*}

\subsection{Spectra Quality Check} \label{sec:spectra}
Metal-poor stars have weak metal lines which make abundance measurements uncertain.  Therefore, we visually examine the spectra of the 333 stars from dwarf galaxies in our sample to assess the quality of the ASPCAP fit for spectral lines of elements we are using.  We use the strongest spectral lines from the line list \citep{apogee_specline} used in the APOGEE abundance pipeline, for a total of 18 lines, with at least 2 lines inspected per element.  Ultimately, we find that Al, O, Mg, Si,  Ti\footnote{Ti is labeled as `deviant' from literature expectations in DR17, however visual inspection shows the lines are well-fit for the stars in our sample.}, and Ni show absorption features, which are generally well-fit by the model. We conclude that it is reasonable to proceed with our analysis using these abundance measurements.  We also include Mn in our analysis, but note that these lines were not as well-fit, and that these abundance measurements should be treated with caution and assumed to be upper limits. Figure \ref{fig:spec_ex} shows a typical example of ASPCAP line fits for the stars and elements included in our sample. In addition to a visual inspection, we implement a quality cut to all abundance measurements whereby we calculate the $\chi^2$ for the ASPCAP best fit model for each line (i.e. $\chi^2 = \Sigma\frac{(\rm{x - y})^2}{\rm{\delta x}^2}$) and remove individual abundance measurements that do not have at least one line with a reduced $\chi^2 > 6$ from consideration in the analysis.

\section{Calibration of APOGEE Abundance Uncertainties} \label{sec:calib}
There are two sources of variability in our abundance measurements that we define: $\delta x$, the internal measurement uncertainty, and $s$, the intrinsic scatter.  The intrinsic scatter of a data set quantifies the "true" scatter among observed stars generated by physical processes. Equation \ref{eq:IS_basic} is a basic calculation of intrinsic scatter that assumes the data are Gaussian with standard deviation $\sigma$ and have the same uncertainties, $\delta x$.  From Equation \ref{eq:IS_basic}, the inverse relationship between intrinsic scatter and measured error is apparent.
\begin{equation} \label{eq:IS_basic}
    s = \sqrt{\sigma^2 - \overline{\delta x}^2}
\end{equation}

We perform our analysis taking into account each star's individual errors using a maximum likelihood approach to estimate the mean abundance of each element, $\Bar{x_i}$, and intrinsic scatter, $s_i$, of the stars in each object given their individual measurement uncertainties, $\delta x_{i,n}$.  The joint probability function \citep{Walker2006,Ness2018,Li2018} is:
\begin{multline}
    \label{eq:IS_mcmc}
    P(x_{i,n}^o | \Bar{x_i},s_i,\delta x_{i,n})  = \\
    \prod_{n=1}^N 1/ \sqrt{2\pi(\delta x^2_{i,n} + s^2_i)} \cdot \rm{exp}(-\frac{(x^o_{i,n} - \Bar{x_i})^2}{2(\delta x^2_{i,n} + s^2_i)})
\end{multline}

We use the log of the probability function to determine the posterior distribution of the mean and intrinsic scatter using \texttt{emcee} \citep{emcee}. Values for the intrinsic scatter are cited as the mean of the posterior distribution with errors that are the 1$\sigma$ standard deviation of the distribution.  All MCMC calculations were required to have a minimum of 3 stars in the sample.

\begin{figure}
    \centering
    \includegraphics[width=\linewidth]{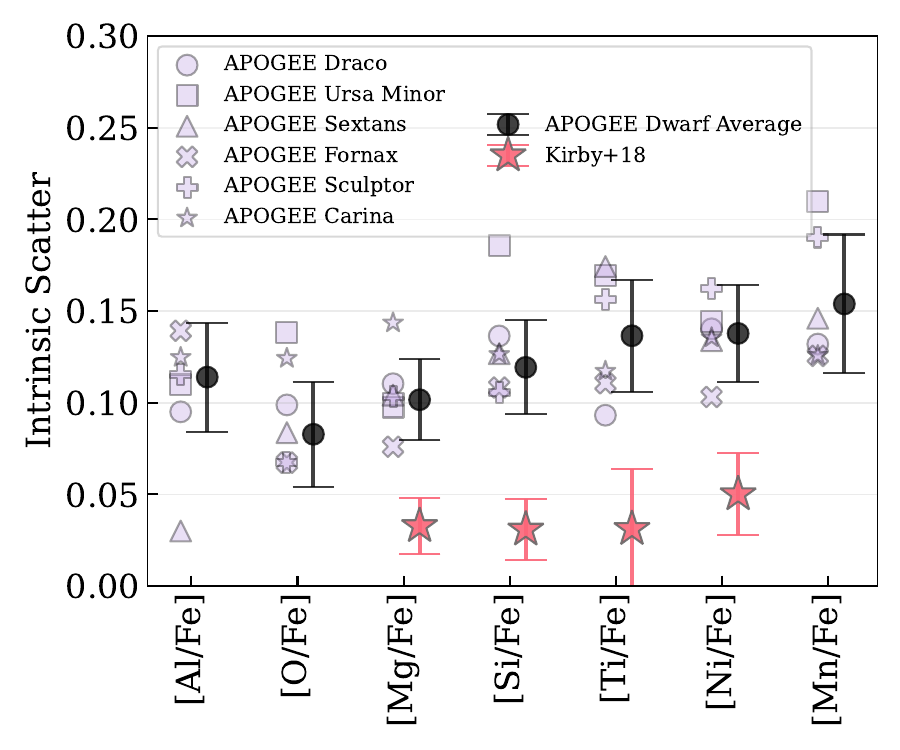}
    \caption{Comparison of the average intrinsic scatters and 1$\sigma$ dispersions of uncalibrated APOGEE dSphs (black) to the average intrinsic scatter of dSphs from \citet{Kirby2018_data} (K18) (pink).  Intrinsic scatter of individual dSphs in APOGEE are in purple. Average APOGEE intrinsic scatters are consistently higher than the \citetalias{Kirby2018_data} intrinsic scatters for elements reported in \citetalias{Kirby2018_data}.}
    \label{fig:IS_dwarf_kirby}
\end{figure}

\subsection{APOGEE Data Intrinsic Scatter}
In Figure \ref{fig:IS_dwarf_kirby} we report intrinsic scatters for individual APOGEE dSphs (purple) using the reported APOGEE abundance uncertainties (henceforth referred to as uncalibrated data).  These range from 0.03 to 0.13 dex for [Al/Fe],  0.07 to 0.14 dex for [O/Fe], 0.08 to 0.14 dex for [Mg/Fe], 0.11 to 0.19 dex for [Si/Fe], 0.09 to 0.17 dex for [Ti/Fe], 0.10 to 0.16 dex for [Ni/Fe], and 0.13 to 0.21 dex for [Mn/Fe].  For each abundance ratio, we also report the average, uncalibrated, intrinsic abundance scatter of all dSphs (in black) and the errorbars are the 1$\sigma$ standard deviation on the mean intrinsic scatters of individual dSphs.  The average scatter ranges from a minimum of 0.08 dex for [O/Fe] to a maximum of 0.15 dex for [Mn/Fe].

APOGEE has errors on the level of intrinsic scatter reported in other literature, therefore we compare the average scatters calculated with the APOGEE data used in our work to average scatters of other data for dwarfs in the literature, namely, the \citet[hereafter \citetalias{Kirby2018_data}]{Kirby2018_data} dataset which carefully accounts for the random and systematic uncertainties.  From this data set, we only consider dSphs with at least 10 observed stars.  This dataset overlaps partially with the APOGEE dataset (Draco, Fornax, Sculptor, Sextans, and Ursa Minor) and includes 3 additional dSphs (Canes Venatici I, Leo I, and Leo II).  Figure \ref{fig:IS_dwarf_kirby} also shows that the uncalibrated data for the dSphs in the APOGEE dataset have higher average intrinsic scatter than the average for those in \citetalias{Kirby2018_data}. This difference is substantial, and ranges from 0.07 to  0.11 dex.

\begin{figure}
    \centering
    \includegraphics[width=\linewidth]{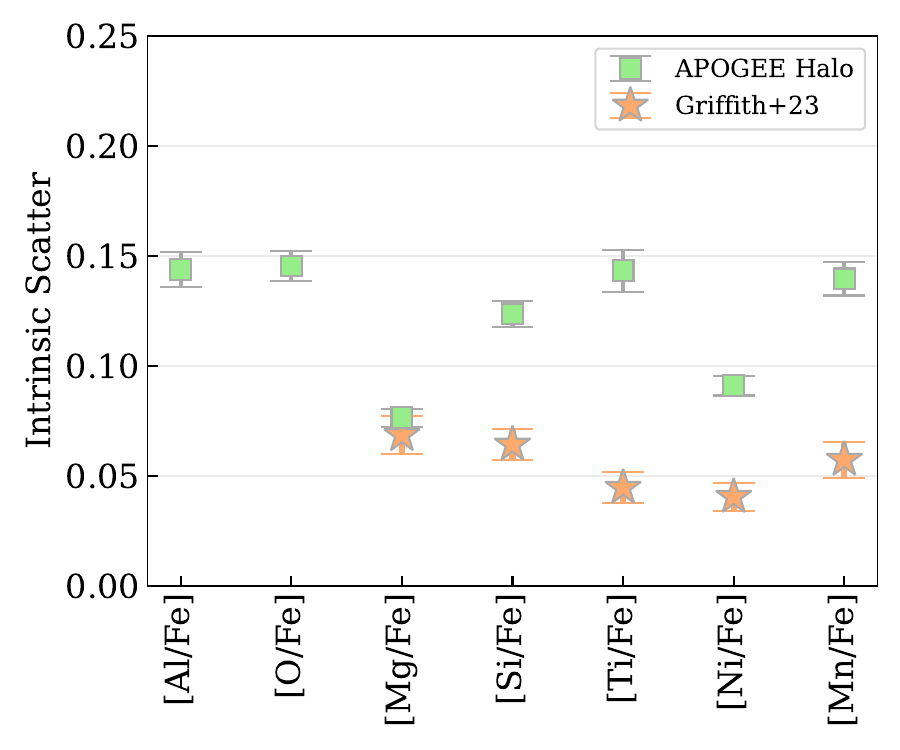}
    \caption{Comparison of intrinsic scatters of uncalibrated APOGEE halo data to intrinsic scatters of data from \citetalias{Griffith23} recalculated using Equation \ref{eq:IS_mcmc}. The recalculated \citetalias{Griffith23} scatters are consistent with those reported in \citetalias{Griffith23}.}
    \label{fig:IS_halo}
\end{figure}

Similarly high intrinsic abundance scatters can be shown for field stars in the APOGEE halo data when compared to intrinsic abundance scatters calculated in the same way from \citetalias{Griffith23} data.  In Figure \ref{fig:IS_halo}, we report the intrinsic abundance scatters in the halo for the same elements. These range from a minimum of 0.08 dex for [Mg/Fe] to a maximum of 0.15 dex for [O/Fe].  We also show that these values range from 0.01 up to 0.1 dex higher than intrinsic scatters for halo stars in \citetalias{Griffith23}. These differences call into question the accuracy of the APOGEE uncertainty estimates for the metal-poor stars.

\subsection{Uncertainty Estimates and Error Calibration}
\subsubsection{Reported APOGEE Uncertainties}
ASPCAP uses FERRE \citep{aspcap}, a $\chi^2$ minimization optimization algorithm, to compare the APOGEE observed spectrum to a grid of synthetic spectra that uses the flux errors computed during data reduction as the weights, and identify the atmospheric parameters and abundances that yield the best fit spectrum.  A first pass of FERRE derives the atmospheric parameters of each star, [$\alpha$/Fe], and [Fe/H], and then a second pass treats the values of these parameters as fixed and derives the abundance measurements.

Uncertainties on the derived parameters are estimated from the inverse of the curvature matrix which depends on the flux error and partial derivatives of the synthetic spectra with the different parameters estimated.  These uncertainties are likely underestimated. Indeed, estimates of abundance scatter in star clusters, yield substantially larger errors than the internal FERRE uncertainty estimates \citep{Holtzman2015}.  These uncertainties were adjusted in recent data releases using repeat observations, but abundance measurements may still have systematic errors for metal-poor stars as the data is sparse in this regime and at low S/N \citep{Jonsson2020}.  Repeat observations provide an empirical estimate of uncertainty as a function of T$\rm{_{eff}}$, [M/H] and S/N (Holtzman et al., in prep).  These empirically derived uncertainties are reported for each abundance in the data release table which we used to calculate the intrinsic scatters in Figures \ref{fig:IS_dwarf_kirby} and \ref{fig:IS_halo}.

As it is unlikely that the intrinsic scatter the APOGEE element abundances are substantially larger than those of \citetalias{Kirby2018_data} as shown in Figure \ref{fig:IS_dwarf_kirby}, we conclude that underestimated errors in APOGEE are the source of inflated $s$ as would be the result of Equation \ref{eq:IS_basic}. The importance of having accurate errors in calculating a meaningful intrinsic scatter motivates us to test and calibrate the uncertainty estimates reported for our stars using an independent approach we develop for this work.

\begin{deluxetable}{c|ccc}[t!]
\tablecaption{Median number of stars in each calibration bin.}
\tablenum{1}
\label{tab:star_count}
\tablehead{
\colhead{}&\colhead{-2.1 < [Fe/H] < -1.55}&\colhead{-1.55 < [Fe/H] < -1}\\
\hline
\multirow{1}{*}{25 < S/N < 125} & 9 & 46
\\
\hline
\multirow{1}{*}{125 < S/N < 225} & 9 & 51
\\
\hline
\multirow{1}{*}{225 < S/N < 325} & 11 & 53
\\
\hline
\multirow{1}{*}{325 < S/N < 425} & 8 & 24
\\
\hline
\multirow{1}{*}{425 < S/N } & 19 & 50
}
\startdata
\enddata
\end{deluxetable}

\subsubsection{High-Fidelity \citetalias{Griffith23} Data Sample}

We use the high-fidelity data presented in \citetalias{Griffith23} to compare to and subsequently calibrate the APOGEE halo element abundance intrinsic scatters.  The \citetalias{Griffith23} sample contains abundance measurements of 12 elements (Na, Mg, Si, Ca, Sc, Ti, V, Cr, Mn, Fe, Co, Ni) for 86 metal-poor ($-2.1 \le \rm{[Fe/H]} \le -1$) subgiants in the solar neighborhood ($d \le 2500 \, \rm{pc}$) observed with the Potsdam Echelle Polarimetric and Spectroscopic Instrument (PEPSI) \citep{PEPSI2015} on the Large Binocular Telescope (LBT).  Spectra were obtained with resolution $R = 50,000$, and $S/N>125$ and $S/N > 236$ respectively in the wavelength ranges of 4260–4800 and 5440–6270 \AA.

The primary goal of the survey was to measure the intrinsic abundance scatter.  To do so, they selected a stellar sample with tight T$\rm{_{eff}}$ and logg constraints to minimize systematic abundance trends and undertook a careful accounting of measurement errors.  \citetalias{Griffith23} assume that their data has negligible differential systematic scatter, and account only for the photon-noise scatter.  Lacking abundances measurements from repeat observations of the same star, they use the error spectrum calculated in the PEPSI reduction pipeline to simulate 10 synthetic spectra for each star.  They vary the flux at each wavelength by an amount drawn from a Gaussian with a standard deviation equivalent to the error.  The new realizations of the spectra are used to recalculate the abundances and the standard deviation is adopted as the photon-noise.  \citetalias{Griffith23} find that the median photon-noise uncertainty is  $\rm{\sigma_{phot,med}}<0.04$ dex. They also consider NLTE effects, but conclude that while the shape of the abundance pattern is influenced by NTLE effects, the abundance scatter is not.

\citetalias{Griffith23} find intrinsic scatters ranging from $0.04-0.08$ dex in [X/Fe] for element abundances in our set (Mg, Si, Ti, Ni, Mn). This is greater than the photon-noise uncertainty by a factor of 2 for most elements.  Overall, because the median photon-noise uncertainties are smaller than the scatter, small inaccuracies in the abundance uncertainty will have little impact on the intrinsic scatter.  Under the assumption that these errors may be underestimated due to the assumption of negligible differential systematic scatter, the method provides robust upper limits on the intrinsic scatter of the element abundances for this population.  If on the other hand, uncertainties on the data were underestimated by, for example, a factor of 2, intrinsic scatters would only increase by 0.01 dex for each element in our sample, well within the uncertainty on the measurement.

\begin{deluxetable*}{cc|ccccc}[t!]
\tablecaption{Average added error terms, $a$, for each S/N and [Fe/H] bin for each abundance.  Abundances for which $a$ was extrapolated are in italics.}
\tablenum{2}
\label{tab:add_err}
\tablehead{
\colhead{} & \colhead{} & \colhead{25 < S/N < 125} & \colhead{125 < S/N < 225} & \colhead{225 < S/N < 325} & \colhead{325 < S/N < 425} & \colhead{425 < S/N}\\
\hline
\multirow{2}{*}{\textit{[Al/Fe], [O/Fe]}}    & -2.1 < [Fe/H] < -1.55 & $0.36 \pm 0.10$ & $0.23 \pm 0.12$ & $0.19 \pm 0.09$ & $0.15 \pm 0.11$ & $0.09 \pm 0.07$\\
                                            & -1.55 < [Fe/H] < -1   & $0.23 \pm 0.03$ & $0.08 \pm 0.04$ & $0.06 \pm 0.06$ & $0.04 \pm 0.02$ & $0.08 \pm 0.04$\\
\hline
\multirow{2}{*}{[Mg/Fe]}    & -2.1 < [Fe/H] < -1.55 & $0.46 \pm 0.15$ & $0.04 \pm 0.03$ & $0.05 \pm 0.04$ & $0.04 \pm 0.03$ & $0.04 \pm 0.03$\\
                            & -1.55 < [Fe/H] < -1   & $0.24 \pm 0.03$ & $0.01 \pm 0.01$ & $0.01 \pm 0.01$ & $0.02 \pm 0.02$ & $0.04 \pm 0.02$\\
\hline
\multirow{2}{*}{[Si/Fe]}    & -2.1 < [Fe/H] < -1.55 & $0.43 \pm 0.13$ & $0.22 \pm 0.08$ & $0.24 \pm 0.07$ & $0.03 \pm 0.03$ & $0.02 \pm 0.02$\\
                            & -1.55 < [Fe/H] < -1   & $0.26 \pm 0.03$ & $0.09 \pm 0.02$ & $0.04 \pm 0.02$ & $0.03 \pm 0.02$ & $0.08 \pm 0.01$\\
\hline
\multirow{2}{*}{[Ti/Fe]}    & -2.1 < [Fe/H] < -1.55 & $0.42 \pm 0.22$ & $0.19 \pm 0.11$ & $0.32 \pm 0.11$ & $0.31 \pm 0.22$ & $0.22 \pm 0.05$\\
                            & -1.55 < [Fe/H] < -1   & $0.20 \pm 0.03$ & $0.13 \pm 0.02$ & $0.06 \pm 0.02$ & $0.05 \pm 0.03$ & $0.14 \pm 0.06$\\
\hline
\multirow{2}{*}{[Ni/Fe]}    & -2.1 < [Fe/H] < -1.55 & $0.17 \pm 0.07$ & $0.36 \pm 0.13$ & $0.16 \pm 0.05$ & $0.15 \pm 0.06$ & $0.05 \pm 0.02$\\
                            & -1.55 < [Fe/H] < -1   & $0.18 \pm 0.02$ & $0.05 \pm 0.01$ & $0.02 \pm 0.01$ & $0.02 \pm 0.01$ & $0.03 \pm 0.01$\\
\hline
\multirow{2}{*}{[Mn/Fe]}    & -2.1 < [Fe/H] < -1.55 & $0.32 \pm 0.11$ & $0.36 \pm 0.18$ & $0.20 \pm 0.07$ & $0.25 \pm 0.13$ & $0.14 \pm 0.04$\\
                            & -1.55 < [Fe/H] < -1   & $0.25 \pm 0.03$ & $0.11 \pm 0.02$ & $0.17 \pm 0.02$ & $0.08 \pm 0.02$ & $0.07 \pm 0.01$
}
\startdata
\enddata
\end{deluxetable*}

To be consistent with the calculated APOGEE intrinsic scatters, we use Equation \ref{eq:IS_mcmc} to calculate the intrinsic scatter of the \citetalias{Griffith23} data for the 86 stars in their sample, showing the result in Figure \ref{fig:IS_halo}. We report a range of intrinsic scatters in [X/Fe] of 0.04 $-$ 0.07 dex for elements that overlap with APOGEE (Mg, Si, Ti, Ni, Mn).  The calculated intrinsic scatters for \citetalias{Griffith23} are within (Mg, Si, Ti) or slightly lower (Ni, Mn) than the reported intrinsic scatters relative to the two- and three-parameter models reported in Table 4 of \citetalias{Griffith23}.

\subsubsection{Calibration of APOGEE Uncertainties}
To accurately calibrate the APOGEE uncertainties, we construct a comparable sample of halo stars from APOGEE as described in Section \ref{sec:data}, and calculate the intrinsic scatter to compare to the \citetalias{Griffith23} intrinsic scatter.  Figure \ref{fig:IS_halo} highlights the discrepancy between the intrinsic scatter of a comparable APOGEE halo sample and the \citetalias{Griffith23} sample. 

Due to the robustness of intrinsic scatter from the \citetalias{Griffith23} data, we take these measurements as the "ground truth" intrinsic scatter of halo field stars and calibrate the uncertainties on the APOGEE data such that the calculated  intrinsic scatter of the APOGEE halo data matches that of \citetalias{Griffith23}.  In practice, we calculate an additional term, $a$, for every abundance and add it in quadrature to the reported uncertainty on abundance measurements in APOGEE.

\begin{multline}
    \label{eq:add_err}
    P(x_{i,n}^o | \Bar{x_i},s_i,\delta x_{i,n})  = \\
    \prod_{n=1}^N 1/ \sqrt{2\pi(\delta x^2_{i,n} + a^2 + s^2_i)} \cdot \rm{exp}(-\frac{(x^o_{i,n} - \Bar{x_i})^2}{2(\delta x^2_{i,n} + a^2 + s^2_i)})
\end{multline}

Using equation \ref{eq:add_err} we fix the intrinsic scatter, $s$, to the calculated value from the \citetalias{Griffith23} data and maximize the likelihood over $a$.  We bin over the APOGEE halo data in [Fe/H] and S/N as we expect that the total uncertainties are primarily dependent on these quantities.  As the morphology of the absorption features may also be influenced by the effective temperature, a better calibration of the errors may be obtained by binning with respect to $\rm{T_{eff}}$, however we are unable to do this robustly with our sample size.  Due to the small number of high S/N stars at low metallicity, we divide our [Fe/H] range into two bins from $-2.1$~dex to $-1.55$~dex and $-1.55$~dex to $-1$~dex.  We then bin the S/N in ranges of 100, from a S/N of 25 up to 425.  The final bin contains all stars with S/N higher than 425 as there are very few stars in this range and S/N beyond this range are thought to be unphysically high due to persistence effects \citep{APOGEEDR13and14}, though the data does still have very high S/N.  We note that finer binning would result in more precise added error terms on the data, but again, we are limited by the APOGEE halo sample size.  The median number of stars in each bin is listed in Table \ref{tab:star_count}.

We extrapolate added error terms for the remaining abundances not in the \citetalias{Griffith23} dataset as all the abundances in our sample have approximately the same average uncertainty, enabling us model the added error terms for abundances for which we calculate $a$ as a Gaussian.  The extrapolated terms are the mean of this Gaussian distribution\footnote{We also tested a uniform model using the maximum and minimum as the bounds on the range.  The difference between models was insignificant.}.  Our results for $a$ are reported in Table \ref{tab:add_err}.

\begin{figure*}[t!]
    \centering
    \includegraphics[width=\textwidth]{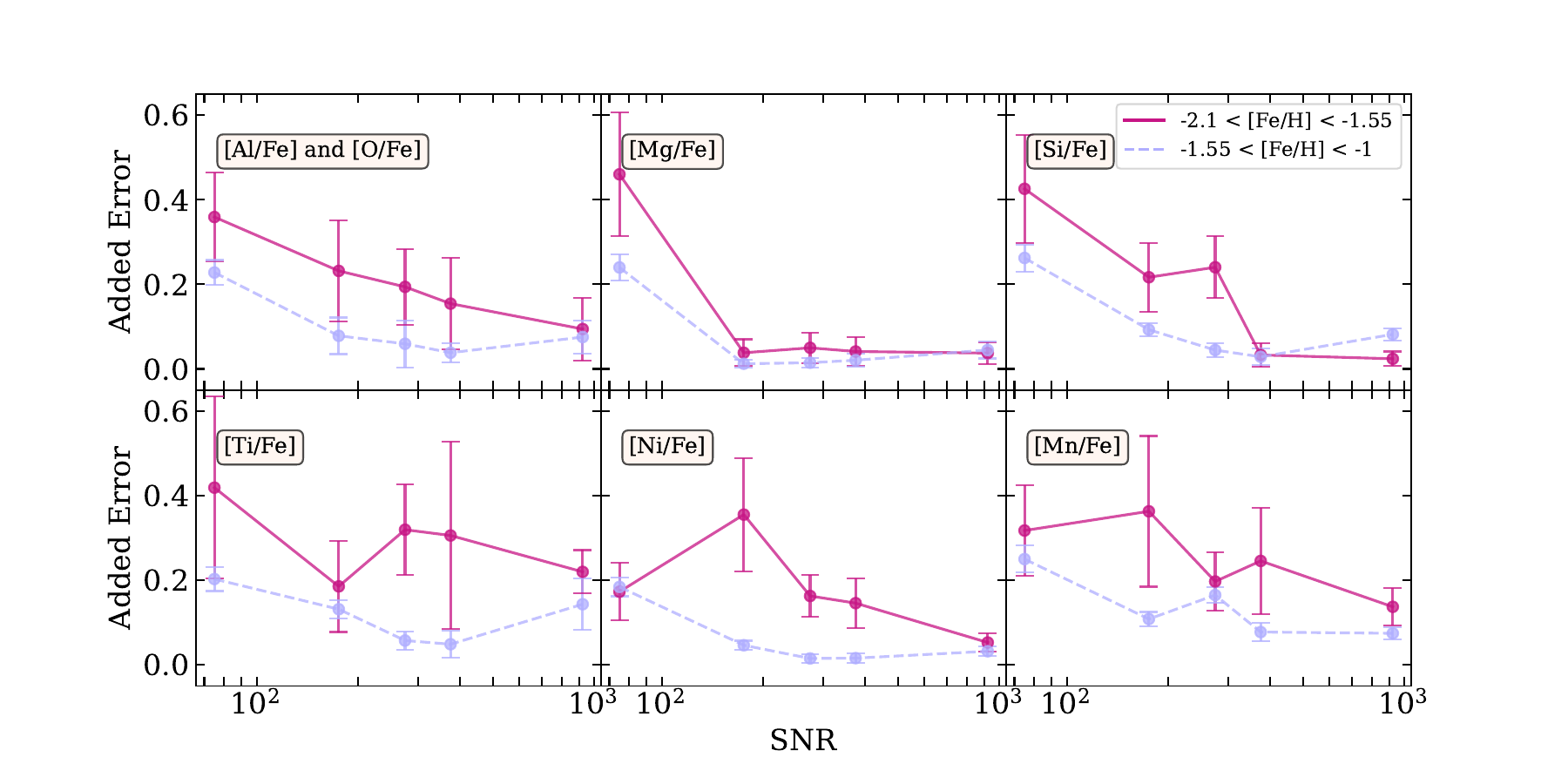}
    \caption{Added error terms for all elemental abundances, binned by S/N and [Fe/H].  Within each elemental abundance, added error terms generally decrease with increasing S/N and [Fe/H].}    \label{fig:add_err}
\end{figure*}

Figure \ref{fig:add_err} shows the added error terms for all abundances explored with the APOGEE data.  There is a clear trend of decreasing added error terms with higher S/N and [Fe/H].  This is expected as higher S/N and higher metallicity data have stronger metal lines. Furthermore, we note that the size of the added error terms at low S/N are on the order of the uncalibrated intrinsic scatter.  This implies the intrinsic scatters are very sensitive to the estimated uncertainty which is a limitation of the data quality.

As the intrinsic scatters reported by \citetalias{Griffith23} are potentially upper limits themselves, the intrinsic scatters measured in this study, calibrated by the added error term, are (while measurements) also potentially upper bounds as they are limited by the accuracy on the uncertainties of the \citetalias{Griffith23} data.  Having constructed the \citetalias{Griffith23} halo and APOGEE samples to be drawn from nearly the same underlying population, this is a robust data-driven method to obtain accurate uncertainties that we can propagate for our analysis of the dSph systems and provides a benchmark for comparison for further studies.

To obtain calibrated uncertainties on the data, the added error terms are applied in quadrature to the reported uncertainty from ASPCAP for each element by drawing from a random Gaussian distribution defined by the posteriors on the added error.  The joint probability is again maximized for the intrinsic scatter as in equation \ref{eq:IS_mcmc}.  Intrinsic scatters for the APOGEE data reported from this point forward use this calibrated data.

\section{Intrinsic Scatter from Calibrated APOGEE Data} \label{sec:scatter}
In this section, we apply the derived added error terms for the individual abundances to the APOGEE data for the Milky Way stellar halo, dSphs, and Milky Way globular clusters, and compare the calibrated and uncalibrated intrinsic scatters.  A table of the calibrated intrinsic scatters as well as object properties is provided in the Appendix.

\subsection{Milky Way Stellar Halo}

\begin{figure}
    \centering
    \includegraphics[width=\linewidth]{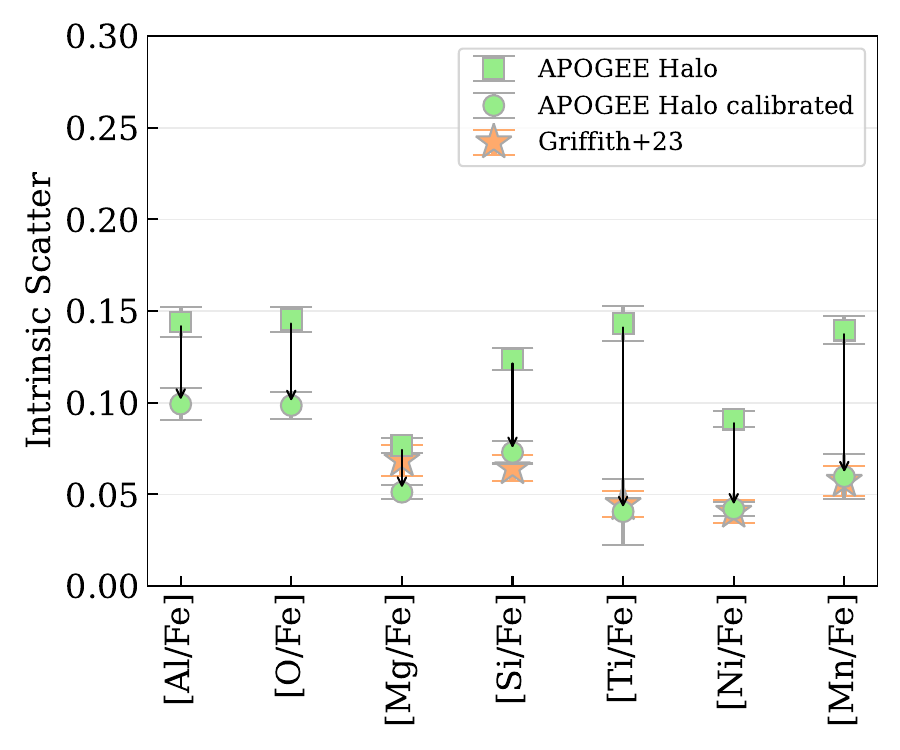}
    \caption{Calibrated intrinsic scatters for the APOGEE halo compared to \citetalias{Griffith23}.  Calibrated APOGEE intrinsic scatters are consistent with the recalculated \citetalias{Griffith23} scatters.}
    \label{fig:a17_halo_IS}
\end{figure}

To validate our calibrated uncertainty estimates, we first compare the calibrated intrinsic scatter of the APOGEE halo sample to the \citetalias{Griffith23} halo sample as shown in Figure \ref{fig:a17_halo_IS}.  Calibrated intrinsic scatters for all abundances decrease by at least 0.02~dex compared to the uncalibrated intrinsic scatters. For those abundances overlapping with the \citetalias{Griffith23} data, intrinsic scatters for the calibrated APOGEE data are approximately the same as the intrinsic scatters derived from the \citetalias{Griffith23} data, as expected.  Specifically, we report intrinsic scatters of 0.07 dex for [Si/Fe], 0.04 dex for [Ti/Fe], 0.04 dex for [Ni/Fe], and 0.06 for [Mn/Fe] which are all within 0.01 dex of \citetalias{Griffith23}.  We report an intrinsic scatter of 0.05 for [Mg/Fe], notably nearly 0.02 dex lower than the \citetalias{Griffith23} value. This suggests that the added error term may be too large for some data and that the data may need to be more finely binned to get an accurate estimate of the added error for each star, which would require a larger sample of low-metallicity stars.  In general, the differences compared to 

\begin{figure}
    \centering
    \includegraphics[width=\linewidth]{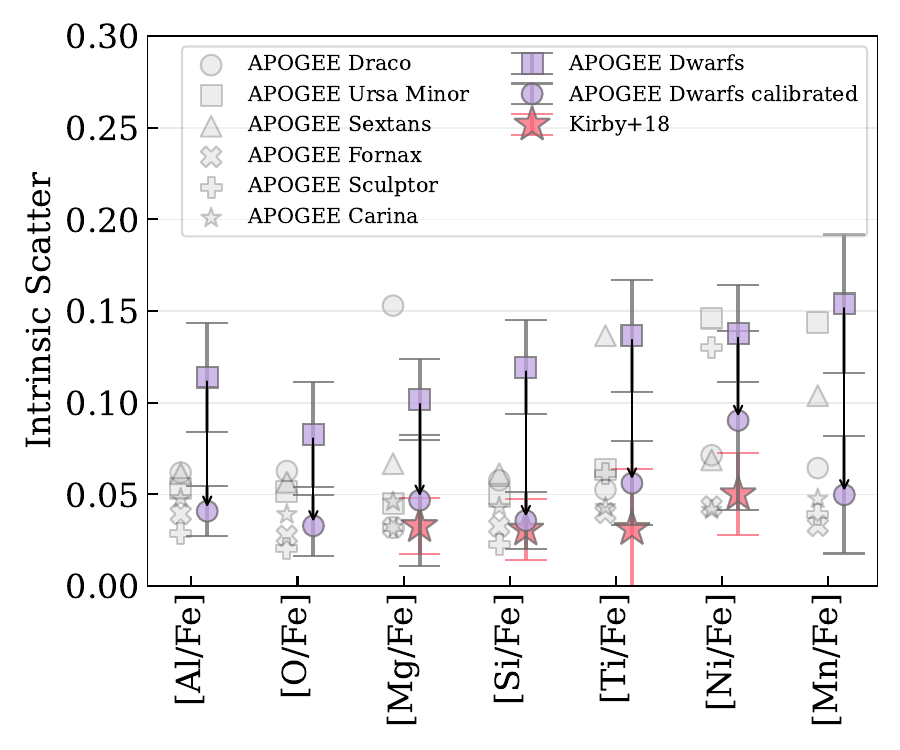}
    \caption{Comparison of average calibrated APOGEE intrinsic scatters for dSphs (purple circles) with average intrinsic scatter from \citetalias{Kirby2018_data} dSphs (pink stars).  Individual calibrated APOGEE dwarfs are in grey.  Purple squares are the average uncalibrated APOGEE intrinsic scatter for dSphs.  The calibrated scatters are generally comparable to the \citetalias{Kirby2018_data} scatters.}
    \label{fig:IS_dwarf_kirby_final}
\end{figure}
\citetalias{Griffith23} are likely due to the coarse binning for the added error term.  In addition, since the maximization of the joint probability function accounts for the individual errors on stars, and the added error terms are not calibrated for individual stars; this likely introduces some small differences between the intrinsic scatter in the \citetalias{Griffith23} data and the calibrated APOGEE data. Nevertheless, these differences are slight and the comparison in Figure \ref{fig:a17_halo_IS} validates our procedure.

\subsection{Milky Way dSphs}

With the robustness of our calibration confirmed, we now turn to an analysis of the dSphs.  Figure \ref{fig:IS_dwarf_kirby_final} shows the comparison between the calibrated and uncalibrated average dSph intrinsic scatter, calibrated intrinsic scatters for individual dSphs in APOGEE, and the average intrinsic scatters from the \citetalias{Kirby2018_data} sample.  The calibrated average intrinsic scatter of dSphs is reduced by up to 0.10 dex over the uncalibrated data, making the average scatter of dSphs in the APOGEE data comparable to the average scatter of dSphs in \citetalias{Kirby2018_data}.

The data also appear to show that the dSphs have lower intrinsic scatter and dispersion on average in the light and $\alpha$-elements (average 0.043 $\pm$ 0.008~dex) than in the Fe-peak elements (average 0.070 $\pm$ 0.027~dex), though this is within statistical uncertainties at the 1$\sigma$ level.  If this variation does exist, it may arise from the difference in time-scales for core-collapse versus Type Ia SNe.  More common core-collapse SNe occur relatively frequently and on short-timescales which could result in greater mixing over time and lower impact from any single event, leading to lower intrinsic scatter.  Conversely, less frequent longer-timescale Type Ia SNe could result in less mixing before star formation episodes, thereby increasing intrinsic scatter.  While we do not see a statistically significant difference in the average intrinsic scatter between the light and $\alpha$-elements, and the Fe-peak elements, future data can test this.

\subsection{Milky Way Globular Clusters}

For comparison with the Milky Way stellar halo and dSphs, we apply our method to the GCs available in APOGEE DR17. These old, gravitationally bound stellar systems have traditionally been thought to form as a single object, which should give them small intrinsic abundance scatter; however, this scenario is made complicated by the presence of multiple stellar populations in most GCs \citep{Bedin2004,Bellini2010,Piotto2012,Piotto2015,Bastian2018,Milone2018,Milone&Marino2022}. Furthermore, these populations are only pronounced in specific elements, typically associated with high-temperature H-burning \citep{Bastian2018}. Of the abundances examined in this work, Al and O are the only ones that shows multiple populations in GCs. 

\begin{figure}
    \centering
    \includegraphics[width=\linewidth]{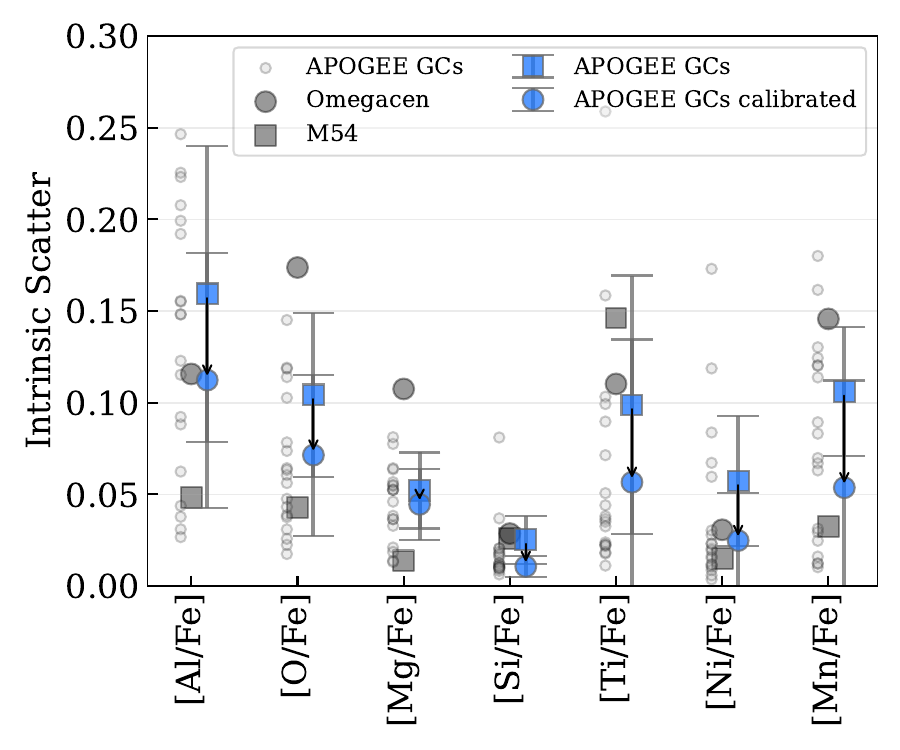}
    \caption{Comparison of average APOGEE intrinsic scatters for globular clusters before (blue squares) and after (blue circles) calibration. Individual objects are in grey.}
    \label{fig:a17_GC_IS}
\end{figure}

Figure \ref{fig:a17_GC_IS} shows the average intrinsic scatters of 1604 stars across 19 GCs (median 79 stars per GC) that were targeted in APOGEE DR17 (see the Appendix for a list of GCs).  We separately analyze Omega Centauri and M54, as they are suspected cores of disrupted dwarf galaxies \citep{Neumayer+2020}. The GCs have a median S/N of 157, and [Fe/H] of -1.1. The average intrinsic scatter is 0.11 dex for [Al/Fe] and 0.07 dex for [O/Fe]; higher than the $\alpha$-elements and Fe-peak elements, each of which have an average scatter of 0.04 dex. This is consistent with the multiple stellar populations mentioned previously.  We also note that our results are consistent with abundance dispersions previously measured in GCs \citep[e.g.][]{Yong2013}.

\citet{Meszaros+2015} presented a similar study of GCs in APOGEE DR10, comparing the observed scatter of the abundances with their uncertainty estimates. Similar to our work, they find significant scatter in [Al/Fe], attributing this to the light-element multiple populations. Furthermore, they highlight that the scatter decreases with increasing [Fe/H]. We find similar trends in [Al/Fe], [Mn/Fe], and possibly [Ni/Fe].

\section{Discussion and Conclusions} \label{sec:conc}
\subsection{Summary}
In this work we introduce a methodology for calibrating abundance uncertainties derived from mid-resolution surveys by leveraging smaller, high-fidelity, high-resolution data sets.  Data-driven calibration or label-transfer efforts have previously been employed to garner greater value from low S/N data \citep[e.g.][]{Casey2017,Ho2017,Rice2020,Xiang2021,Xiang2022}. However, these typically rely on stars in common between the surveys. Our approach does not rely on this, but instead utilizes population statistics. We apply our method to calibrate the measurement errors and calculate the intrinsic abundance scatter of metal-poor stars for elements Al, O, Mg, Si, Ti, Ni, and Mn in dSphs, GCs, and the Milky Way stellar halo.  Our calculation relies on the calibration of the reported APOGEE abundance uncertainties and is specifically for APOGEE DR17 metal-poor red giant stars ($-2.1 \le $[Fe/H]$ \le -1$) and in the temperature range $4000 \, \rm{K} \, \le T \le \, 5000 \, \rm{K}$.

Making measurements of the intrinsic abundance scatter of dSphs, GCs, and the halo with the APOGEE data, particularly for metal-poor stars, is important for providing constraints for simulations and linking to physical drivers.  Doing this requires using data at lower metallicities and S/N than the majority of APOGEE spectra, in a regime where there is little previous work. As the magnitude of the uncertainties impact the measurement of intrinsic scatter, it is vital to test the accuracy of the APOGEE reported uncertainty. In doing so, we found them to be underestimated and using the method introduced in this paper, we derive the complete uncertainties on the APOGEE data in this regime.

To do this, we use a high fidelity reference set of halo stars (\citetalias{Griffith23}) to provide ground truth information about the intrinsic scatter of element abundances in the stellar halo. We construct a comparable sample of field halo stars from the APOGEE data using the same distance and metallicity criteria as \citetalias{Griffith23}.  Our primary assumption in this work is that the underlying population of stars in the reference set (\citetalias{Griffith23}) is comparable to the halo population in the calibration set we use from APOGEE.  Comparing the intrinsic scatters between the two datasets, we find that the intrinsic scatters of the uncalibrated APOGEE halo data are substantially higher than \citetalias{Griffith23}, indicating the APOGEE uncertainties are underestimated.  This is unsurprising as the reference objects used in APOGEE for providing empirical uncertainty estimates are sparse in the metal-poor and low S/N regime (S/N $\sim$ 50) \citep{Jonsson2020}.  We note that the \citetalias{Griffith23} uncertainties themselves are very small so their intrinsic scatters, which we take as ground truth, are minimally impacted by any inaccuracy in these. However, strictly, any underestimation in the \citetalias{Griffith23} uncertainties would result in lower intrinsic scatter, and so these data set an upper limit on the ground truth intrinsic scatter and therefore a lower limit on the complete uncertainties for APOGEE.

To calibrate the APOGEE uncertainties, we determine the uncertainty on the APOGEE halo data required to match the intrinsic abundance scatter of the halo population from \citetalias{Griffith23}. We subsequently calculate an added error term for every abundance as a function of [Fe/H] binned across a range of -2.1 to -1 and S/N binned across a range of 25 to 1420, which we can apply to stars in this parameter space. 

After calibrating the uncertainties for each of the stars in the dSphs, we find that on average, dSphs have comparatively low intrinsic scatter in the light and $\alpha$-elements (on average, 0.04 dex)  and higher intrinsic scatter in the Fe-peak elements (on average, 0.07 dex).  We also undertake a similar analysis for GCs, but show that in contrast to the dSphs, GCs have high intrinsic scatter in the light elements (Al and O with 0.11 and 0.07 dex respectively), and no clear pattern in intrinsic scatter among the $\alpha$- and Fe-peak elements. This may suggest that for elements ejected through higher-energy processes, mixing is most efficient on small scales.

\subsection{Comparison of dSphs, MW Halo, and GCs}
\begin{figure}[t!]
    \centering
    \includegraphics[width=\linewidth]{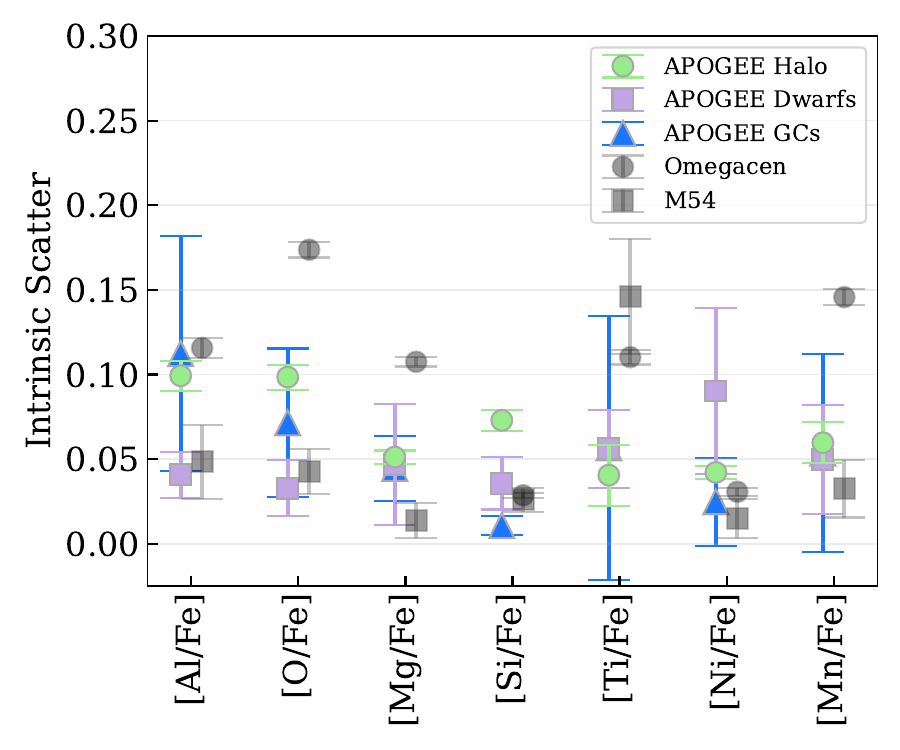}
    \caption{Comparison of the average and dispersion of the observed intrinsic scatters between the stellar halo, dSphs, and GCs.}
    \label{fig:IS_comp_a17}
\end{figure}

Comparing the average intrinsic abundance scatters of dSphs, GCs, and the stellar halo enables us to set constraints on the origin of scatter and the formation histories of different structures.  Figure \ref{fig:IS_comp_a17} shows that the intrinsic scatter of the halo is more similar to the average scatter of GCs than it is to the average scatter of dSphs.  However, this is not to necessarily say that the halo and GCs are formed of the same populations of stars, rather, that we distinctly see the signatures of multiple populations in both.  This may be an indicator of the existence of stars from destroyed GCs in the halo, a theory supported by the high abundance dispersions in Al, O, and Si over Fe found in the in situ halo \citep{Belokurov2022} which have been linked to an increase contribution from GCs \citep{Belokurov2023}.  This may also be a signature of the combination of dwarfs with multiple populations of different element abundance means and scatters, as we expect for the composition of the halo, which would contribute to a larger scatter.

Additionally, that the classical dSphs have a different chemical composition than the halo has been shown both in observational \citep{Unavane1996, Venn2004} and simulated data \citep{Robertson2005,Font2006,Johnston2008,Cunningham2022,HortaCunningham2023}.  Much of the difference is attributed to the short star formation histories of accreted dwarfs compared to longer-lived classical dSphs. The higher scatter seen in the halo could be interpreted as a result of early quenching of star formation in dwarfs destroyed by the halo, where metals do not have time to mix and end up in stars before star formation ceases.  In comparison, the classical dSphs, such as those in our sample have more extended star formation histories which allow for more complete mixing, resulting in lower intrinsic scatters.

\subsection{Future Prospects for Observations and Simulations}
The ability to account for the complete uncertainties in the metal-poor regime of survey data is critical in order to fully exploit these data. Large-scale surveys also offer unique coverage of wider areas of dSphs than smaller studies, which typically focus on the core.  Broader area coverage in general enables studies of element abundance scatter across the Galaxy, as well as spatially dependent scatter.  Currently, available survey data of dSphs stars are generally low S/N ($< 70$ median in APOGEE), are limited to the brightest stars, and element abundance measurement uncertainties are nearly as high as the total abundance scatter.  In order to make high-fidelity measurements of the intrinsic scatter of element abundances of low-metallicity dSph systems, reasonable sample sizes of high resolution (e.g. R $>$ 50,000), high S/N spectra (e.g. $>$ 100) are required.

However, our efforts to calibrate large medium-resolution survey data using small high resolution, high S/N studies makes the most use of the full parameter space of the available data, as well as provides a better understanding of the unaccounted uncertainties in the metal-poor regime in APOGEE, which is relevant to an ensemble of studies. Our results subsequently serve as a baseline for the next generation of data. High-resolution, high S/N spectra can be obtained with upcoming 30-meter class telescope programs such as the Giant Magellan Telescope (GMT), European Extremely Large Telescope (E-ELT), and Thirty Meter Telescope (TMT).  Additionally, 30-meter class telescopes are expected to reach up to the 25th magnitude in the V-band, or 2 magnitudes deeper than existing observations of resolved stars in dSphs \citep{Ji2019}.   The intrinsic scatter in element abundances as we showcase here can serve as an important test and comparison for future ELT studies where measurement uncertainties will undoubtedly be smaller, and will serve as another source of calibration data for the uncertainty estimates used in this work.

Intrinsic abundance scatter of dSphs and other galactic systems also provides important constraints on theoretical simulations and models for studying the physical origins of abundance scatter and formation histories of dwarf galaxies, halos, and star clusters. In particular, it constrains the amount of gas mixing, the stochasticity of star formation, the variability in and ratio between different nucleosynthetic sources, and the chemical signatures produced by the first stars.  A better understanding of all of these processes can assist us in interpreting the element abundance scatter we see in metal-poor stars today. An in-depth analysis of high-resolution simulations will help us to interpret the trends we have found in the APOGEE data. 

Of particular interest are star-by-star models (e.g. \texttt{\AE os}, Brauer et al., in prep.; \texttt{GRIFFIN}, \citealt{Lahen+2020}; \texttt{INFERNO}, \citealt{Andersson+2023}; \texttt{LYRA}, \citealt{Gutcke+2021}), which allow for detailed chemical yield models linked to specific stellar types \citep[e.g.][]{Emerick2020,Emerick+2020,Lahen+2023}. The details of when and where stars inject enriched material likely plays a role for abundance features like the intrinsic scatter. Furthermore, star-by-star models sample an IMF as opposed to treating stars as unresolved mono-age populations.  This latter feature is important in modeling the early universe to resolve low-mass stellar systems due to the formation of only handfuls of massive stars that inject enriched material into each halo. Future work with simulations, including analyzing how supernovae ejecta mix in the interstellar and intergalactic mediums both spatially and temporally, and how the intrinsic abundance scatters of stars in each halo evolves as these elements mix, will be anchored by the measurements we provide.
\bigskip
\section{Acknowledgments}
The authors thank Professor Alexander Ji for extensive discussions on methodology for calculating intrinsic scatter with the ASPCAP measurements of APOGEE data, Professors Mary Putman and Kathryn Johnston, and Dr. Emily Cunningham for discussions about early results, and Drs. Christian Hayes and Rachael Beaton for their time and insight on the usability of dwarf galaxy abundances at low metallicities.  The authors also thank the referee for a detailed and insightful report.

J.M. acknowledges support from the National Science Foundation Graduate Research Fellowship under grant DGE-2036197. 
 E.A.\ acknowledges support from the US National Science Foundation under grants AST18-15461 and AST23-07950.  E.J.G. is supported by an NSF Astronomy and Astrophysics Postdoctoral Fellowship under award AST-2202135.

\bibliographystyle{yahapj}
\bibliography{refs}

\begin{thebibliography}{}
\providecommand\natexlab[1]{#1}
\providecommand\JournalTitle[1]{#1}

\bibitem[{{Abdurro'uf} {et~al.}(2022){Abdurro'uf}, {Accetta}, {Aerts}, {Silva Aguirre}, {Ahumada}, {Ajgaonkar}, {Filiz Ak}, {Alam}, {Allende Prieto}, {Almeida}, {Anders}, {Anderson}, {Andrews}, {Anguiano}, {Aquino-Ort{\'\i}z}, {Arag{\'o}n-Salamanca}, {Argudo-Fern{\'a}ndez}, {Ata}, {Aubert}, {Avila-Reese}, {Badenes}, {Barb{\'a}}, {Barger}, {Barrera-Ballesteros}, {Beaton}, {Beers}, {Belfiore}, {Bender}, {Bernardi}, {Bershady}, {Beutler}, {Bidin}, {Bird}, {Bizyaev}, {Blanc}, {Blanton}, {Boardman}, {Bolton}, {Boquien}, {Borissova}, {Bovy}, {Brandt}, {Brown}, {Brownstein}, {Brusa}, {Buchner}, {Bundy}, {Burchett}, {Bureau}, {Burgasser}, {Cabang}, {Campbell}, {Cappellari}, {Carlberg}, {Wanderley}, {Carrera}, {Cash}, {Chen}, {Chen}, {Cherinka}, {Chiappini}, {Choi}, {Chojnowski}, {Chung}, {Clerc}, {Cohen}, {Comerford}, {Comparat}, {da Costa}, {Covey}, {Crane}, {Cruz-Gonzalez}, {Culhane}, {Cunha}, {Dai}, {Damke}, {Darling}, {Davidson}, {Davies}, {Dawson}, {De Lee}, {Diamond-Stanic}, {Cano-D{\'\i}az}, {S{\'a}nchez},
  {Donor}, {Duckworth}, {Dwelly}, {Eisenstein}, {Elsworth}, {Emsellem}, {Eracleous}, {Escoffier}, {Fan}, {Farr}, {Feng}, {Fern{\'a}ndez-Trincado}, {Feuillet}, {Filipp}, {Fillingham}, {Frinchaboy}, {Fromenteau}, {Galbany}, {Garc{\'\i}a}, {Garc{\'\i}a-Hern{\'a}ndez}, {Ge}, {Geisler}, {Gelfand}, {G{\'e}ron}, {Gibson}, {Goddy}, {Godoy-Rivera}, {Grabowski}, {Green}, {Greener}, {Grier}, {Griffith}, {Guo}, {Guy}, {Hadjara}, {Harding}, {Hasselquist}, {Hayes}, {Hearty}, {Hern{\'a}ndez}, {Hill}, {Hogg}, {Holtzman}, {Horta}, {Hsieh}, {Hsu}, {Hsu}, {Huber}, {Huertas-Company}, {Hutchinson}, {Hwang}, {Ibarra-Medel}, {Chitham}, {Ilha}, {Imig}, {Jaekle}, {Jayasinghe}, {Ji}, {Johnson}, {Jones}, {J{\"o}nsson}, {Katkov}, {Khalatyan}, {Kinemuchi}, {Kisku}, {Knapen}, {Kneib}, {Kollmeier}, {Kong}, {Kounkel}, {Kreckel}, {Krishnarao}, {Lacerna}, {Lane}, {Langgin}, {Lavender}, {Law}, {Lazarz}, {Leung}, {Leung}, {Lewis}, {Li}, {Li}, {Lian}, {Liang}, {Lin}, {Lin}, {Lin}, {Lintott}, {Long}, {Longa-Pe{\~n}a}, {L{\'o}pez-Cob{\'a}}, {Lu},
  {Lundgren}, {Luo}, {Mackereth}, {de la Macorra}, {Mahadevan}, {Majewski}, {Manchado}, {Mandeville}, {Maraston}, {Margalef-Bentabol}, {Masseron}, {Masters}, {Mathur}, {McDermid}, {Mckay}, {Merloni}, {Merrifield}, {Meszaros}, {Miglio}, {Di Mille}, {Minniti}, {Minsley}, {Monachesi}, {Moon}, {Mosser}, {Mulchaey}, {Muna}, {Mu{\~n}oz}, {Myers}, {Myers}, {Nadathur}, {Nair}, {Nandra}, {Neumann}, {Newman}, {Nidever}, {Nikakhtar}, {Nitschelm}, {O'Connell}, {Garma-Oehmichen}, {Luan Souza de Oliveira}, {Olney}, {Oravetz}, {Ortigoza-Urdaneta}, {Osorio}, {Otter}, {Pace}, {Padilla}, {Pan}, {Pan}, {Parikh}, {Parker}, {Peirani}, {Pe{\~n}a Ram{\'\i}rez}, {Penny}, {Percival}, {Perez-Fournon}, {Pinsonneault}, {Poidevin}, {Poovelil}, {Price-Whelan}, {B{\'a}rbara de Andrade Queiroz}, {Raddick}, {Ray}, {Rembold}, {Riddle}, {Riffel}, {Riffel}, {Rix}, {Robin}, {Rodr{\'\i}guez-Puebla}, {Roman-Lopes}, {Rom{\'a}n-Z{\'u}{\~n}iga}, {Rose}, {Ross}, {Rossi}, {Rubin}, {Salvato}, {S{\'a}nchez}, {S{\'a}nchez-Gallego}, {Sanderson}, {Santana
  Rojas}, {Sarceno}, {Sarmiento}, {Sayres}, {Sazonova}, {Schaefer}, {Schiavon}, {Schlegel}, {Schneider}, {Schultheis}, {Schwope}, {Serenelli}, {Serna}, {Shao}, {Shapiro}, {Sharma}, {Shen}, {Shetrone}, {Shu}, {Simon}, {Skrutskie}, {Smethurst}, {Smith}, {Sobeck}, {Spoo}, {Sprague}, {Stark}, {Stassun}, {Steinmetz}, {Stello}, {Stone-Martinez}, {Storchi-Bergmann}, {Stringfellow}, {Stutz}, {Su}, {Taghizadeh-Popp}, {Talbot}, {Tayar}, {Telles}, {Teske}, {Thakar}, {Theissen}, {Tkachenko}, {Thomas}, {Tojeiro}, {Hernandez Toledo}, {Troup}, {Trump}, {Trussler}, {Turner}, {Tuttle}, {Unda-Sanzana}, {V{\'a}zquez-Mata}, {Valentini}, {Valenzuela}, {Vargas-Gonz{\'a}lez}, {Vargas-Maga{\~n}a}, {Alfaro}, {Villanova}, {Vincenzo}, {Wake}, {Warfield}, {Washington}, {Weaver}, {Weijmans}, {Weinberg}, {Weiss}, {Westfall}, {Wild}, {Wilde}, {Wilson}, {Wilson}, {Wilson}, {Wolf}, {Wood-Vasey}, {Yan}, {Zamora}, {Zasowski}, {Zhang}, {Zhao}, {Zheng}, {Zheng}, \& {Zhu}}]{SDSSDR17}
{Abdurro'uf}, {Accetta}, K., {Aerts}, C., {et~al.} 2022, \href{http://dx.doi.org/10.3847/1538-4365/ac4414}{\JournalTitle{\apjs}, 259, 35}

\bibitem[{{Andersson} {et~al.}(2023){Andersson}, {Agertz}, {Renaud}, \& {Teyssier}}]{Andersson+2023}
{Andersson}, E.~P., {Agertz}, O., {Renaud}, F., \& {Teyssier}, R. 2023, \href{http://dx.doi.org/10.1093/mnras/stad692}{\JournalTitle{\mnras}, 521, 2196}

\bibitem[{{Bastian} \& {Lardo}(2018)}]{Bastian2018}
{Bastian}, N., \& {Lardo}, C. 2018, \href{http://dx.doi.org/10.1146/annurev-astro-081817-051839}{\JournalTitle{\araa}, 56, 83}

\bibitem[{{Bedin} {et~al.}(2004){Bedin}, {Piotto}, {Anderson}, {Cassisi}, {King}, {Momany}, \& {Carraro}}]{Bedin2004}
{Bedin}, L.~R., {Piotto}, G., {Anderson}, J., {et~al.} 2004, \href{http://dx.doi.org/10.1086/420847}{\JournalTitle{\apjl}, 605, L125}

\bibitem[{{Bellini} {et~al.}(2010){Bellini}, {Bedin}, {Piotto}, {Milone}, {Marino}, \& {Villanova}}]{Bellini2010}
{Bellini}, A., {Bedin}, L.~R., {Piotto}, G., {et~al.} 2010, \href{http://dx.doi.org/10.1088/0004-6256/140/2/631}{\JournalTitle{\aj}, 140, 631}

\bibitem[{{Belokurov} \& {Kravtsov}(2022)}]{Belokurov2022}
{Belokurov}, V., \& {Kravtsov}, A. 2022, \href{http://dx.doi.org/10.1093/mnras/stac1267}{\JournalTitle{\mnras}, 514, 689}

\bibitem[{{Belokurov} \& {Kravtsov}(2023)}]{Belokurov2023}
---. 2023, \href{http://dx.doi.org/10.1093/mnras/stad2241}{\JournalTitle{\mnras}, 525, 4456}

\bibitem[{{Bertran de Lis} {et~al.}(2016){Bertran de Lis}, {Allende Prieto}, {Majewski}, {Schiavon}, {Holtzman}, {Shetrone}, {Carrera}, {Garc{\'\i}a P{\'e}rez}, {M{\'e}sz{\'a}ros}, {Frinchaboy}, {Hearty}, {Nidever}, {Zasowski}, \& {Ge}}]{Bertran_de_Lis2016}
{Bertran de Lis}, S., {Allende Prieto}, C., {Majewski}, S.~R., {et~al.} 2016, \href{http://dx.doi.org/10.1051/0004-6361/201527827}{\JournalTitle{\aap}, 590, A74}

\bibitem[{{Blanton} {et~al.}(2017){Blanton}, {Bershady}, {Abolfathi}, {Albareti}, {Allende Prieto}, {Almeida}, {Alonso-Garc{\'\i}a}, {Anders}, {Anderson}, {Andrews}, {Aquino-Ort{\'\i}z}, {Arag{\'o}n-Salamanca}, {Argudo-Fern{\'a}ndez}, {Armengaud}, {Aubourg}, {Avila-Reese}, {Badenes}, {Bailey}, {Barger}, {Barrera-Ballesteros}, {Bartosz}, {Bates}, {Baumgarten}, {Bautista}, {Beaton}, {Beers}, {Belfiore}, {Bender}, {Berlind}, {Bernardi}, {Beutler}, {Bird}, {Bizyaev}, {Blanc}, {Blomqvist}, {Bolton}, {Boquien}, {Borissova}, {van den Bosch}, {Bovy}, {Brandt}, {Brinkmann}, {Brownstein}, {Bundy}, {Burgasser}, {Burtin}, {Busca}, {Cappellari}, {Delgado Carigi}, {Carlberg}, {Carnero Rosell}, {Carrera}, {Chanover}, {Cherinka}, {Cheung}, {G{\'o}mez Maqueo Chew}, {Chiappini}, {Choi}, {Chojnowski}, {Chuang}, {Chung}, {Cirolini}, {Clerc}, {Cohen}, {Comparat}, {da Costa}, {Cousinou}, {Covey}, {Crane}, {Croft}, {Cruz-Gonzalez}, {Garrido Cuadra}, {Cunha}, {Damke}, {Darling}, {Davies}, {Dawson}, {de la Macorra}, {Dell'Agli}, {De
  Lee}, {Delubac}, {Di Mille}, {Diamond-Stanic}, {Cano-D{\'\i}az}, {Donor}, {Downes}, {Drory}, {du Mas des Bourboux}, {Duckworth}, {Dwelly}, {Dyer}, {Ebelke}, {Eigenbrot}, {Eisenstein}, {Emsellem}, {Eracleous}, {Escoffier}, {Evans}, {Fan}, {Fern{\'a}ndez-Alvar}, {Fernandez-Trincado}, {Feuillet}, {Finoguenov}, {Fleming}, {Font-Ribera}, {Fredrickson}, {Freischlad}, {Frinchaboy}, {Fuentes}, {Galbany}, {Garcia-Dias}, {Garc{\'\i}a-Hern{\'a}ndez}, {Gaulme}, {Geisler}, {Gelfand}, {Gil-Mar{\'\i}n}, {Gillespie}, {Goddard}, {Gonzalez-Perez}, {Grabowski}, {Green}, {Grier}, {Gunn}, {Guo}, {Guy}, {Hagen}, {Hahn}, {Hall}, {Harding}, {Hasselquist}, {Hawley}, {Hearty}, {Gonzalez Hern{\'a}ndez}, {Ho}, {Hogg}, {Holley-Bockelmann}, {Holtzman}, {Holzer}, {Huehnerhoff}, {Hutchinson}, {Hwang}, {Ibarra-Medel}, {da Silva Ilha}, {Ivans}, {Ivory}, {Jackson}, {Jensen}, {Johnson}, {Jones}, {J{\"o}nsson}, {Jullo}, {Kamble}, {Kinemuchi}, {Kirkby}, {Kitaura}, {Klaene}, {Knapp}, {Kneib}, {Kollmeier}, {Lacerna}, {Lane}, {Lang}, {Law},
  {Lazarz}, {Lee}, {Le Goff}, {Liang}, {Li}, {Li}, {Lian}, {Lima}, {Lin}, {Lin}, {Bertran de Lis}, {Liu}, {de Icaza Lizaola}, {Long}, {Lucatello}, {Lundgren}, {MacDonald}, {Deconto Machado}, {MacLeod}, {Mahadevan}, {Geimba Maia}, {Maiolino}, {Majewski}, {Malanushenko}, {Malanushenko}, {Manchado}, {Mao}, {Maraston}, {Marques-Chaves}, {Masseron}, {Masters}, {McBride}, {McDermid}, {McGrath}, {McGreer}, {Medina Pe{\~n}a}, {Melendez}, {Merloni}, {Merrifield}, {Meszaros}, {Meza}, {Minchev}, {Minniti}, {Miyaji}, {More}, {Mulchaey}, {M{\"u}ller-S{\'a}nchez}, {Muna}, {Munoz}, {Myers}, {Nair}, {Nandra}, {Correa do Nascimento}, {Negrete}, {Ness}, {Newman}, {Nichol}, {Nidever}, {Nitschelm}, {Ntelis}, {O'Connell}, {Oelkers}, {Oravetz}, {Oravetz}, {Pace}, {Padilla}, {Palanque-Delabrouille}, {Alonso Palicio}, {Pan}, {Parejko}, {Parikh}, {P{\^a}ris}, {Park}, {Patten}, {Peirani}, {Pellejero-Ibanez}, {Penny}, {Percival}, {Perez-Fournon}, {Petitjean}, {Pieri}, {Pinsonneault}, {Pisani}, {Poleski}, {Prada}, {Prakash}, {Queiroz},
  {Raddick}, {Raichoor}, {Barboza Rembold}, {Richstein}, {Riffel}, {Riffel}, {Rix}, {Robin}, {Rockosi}, {Rodr{\'\i}guez-Torres}, {Roman-Lopes}, {Rom{\'a}n-Z{\'u}{\~n}iga}, {Rosado}, {Ross}, {Rossi}, {Ruan}, {Ruggeri}, {Rykoff}, {Salazar-Albornoz}, {Salvato}, {S{\'a}nchez}, {Aguado}, {S{\'a}nchez-Gallego}, {Santana}, {Santiago}, {Sayres}, {Schiavon}, {da Silva Schimoia}, {Schlafly}, {Schlegel}, {Schneider}, {Schultheis}, {Schuster}, {Schwope}, {Seo}, {Shao}, {Shen}, {Shetrone}, {Shull}, {Simon}, {Skinner}, {Skrutskie}, {Slosar}, {Smith}, {Sobeck}, {Sobreira}, {Somers}, {Souto}, {Stark}, {Stassun}, {Stauffer}, {Steinmetz}, {Storchi-Bergmann}, {Streblyanska}, {Stringfellow}, {Su{\'a}rez}, {Sun}, {Suzuki}, {Szigeti}, {Taghizadeh-Popp}, {Tang}, {Tao}, {Tayar}, {Tembe}, {Teske}, {Thakar}, {Thomas}, {Thompson}, {Tinker}, {Tissera}, {Tojeiro}, {Hernandez Toledo}, {de la Torre}, {Tremonti}, {Troup}, {Valenzuela}, {Martinez Valpuesta}, {Vargas-Gonz{\'a}lez}, {Vargas-Maga{\~n}a}, {Vazquez}, {Villanova}, {Vivek}, {Vogt},
  {Wake}, {Walterbos}, {Wang}, {Weaver}, {Weijmans}, {Weinberg}, {Westfall}, {Whelan}, {Wild}, {Wilson}, {Wood-Vasey}, {Wylezalek}, {Xiao}, {Yan}, {Yang}, {Ybarra}, {Y{\`e}che}, {Zakamska}, {Zamora}, {Zarrouk}, {Zasowski}, {Zhang}, {Zhao}, {Zheng}, {Zheng}, {Zhou}, {Zhou}, {Zhu}, {Zoccali}, \& {Zou}}]{SDSS}
{Blanton}, M.~R., {Bershady}, M.~A., {Abolfathi}, B., {et~al.} 2017, \href{http://dx.doi.org/10.3847/1538-3881/aa7567}{\JournalTitle{\aj}, 154, 28}

\bibitem[{{Bovy}(2016)}]{Bovy2016}
{Bovy}, J. 2016, \href{http://dx.doi.org/10.3847/0004-637X/817/1/49}{\JournalTitle{\apj}, 817, 49}

\bibitem[{{Busso} {et~al.}(1999){Busso}, {Gallino}, \& {Wasserburg}}]{Busso1999}
{Busso}, M., {Gallino}, R., \& {Wasserburg}, G.~J. 1999, \href{http://dx.doi.org/10.1146/annurev.astro.37.1.239}{\JournalTitle{\araa}, 37, 239}

\bibitem[{{Casey} {et~al.}(2017){Casey}, {Hawkins}, {Hogg}, {Ness}, {Rix}, {Kordopatis}, {Kunder}, {Steinmetz}, {Koposov}, {Enke}, {Sanders}, {Gilmore}, {Zwitter}, {Freeman}, {Casagrande}, {Matijevi{\v{c}}}, {Seabroke}, {Bienaym{\'e}}, {Bland-Hawthorn}, {Gibson}, {Grebel}, {Helmi}, {Munari}, {Navarro}, {Reid}, {Siebert}, \& {Wyse}}]{Casey2017}
{Casey}, A.~R., {Hawkins}, K., {Hogg}, D.~W., {et~al.} 2017, \href{http://dx.doi.org/10.3847/1538-4357/aa69c2}{\JournalTitle{\apj}, 840, 59}

\bibitem[{{Cui} {et~al.}(2012){Cui}, {Zhao}, {Chu}, {Li}, {Li}, {Zhang}, {Su}, {Yao}, {Wang}, {Xing}, {Li}, {Zhu}, {Wang}, {Gu}, {Luo}, {Xu}, {Zhang}, {Liu}, {Zhang}, {Yang}, {Cao}, {Chen}, {Chen}, {Chen}, {Chen}, {Chu}, {Feng}, {Gong}, {Hou}, {Hu}, {Hu}, {Hu}, {Jia}, {Jiang}, {Jiang}, {Jiang}, {Jin}, {Li}, {Li}, {Li}, {Liu}, {Liu}, {Lu}, {Mao}, {Men}, {Qi}, {Qi}, {Shi}, {Tang}, {Tao}, {Wang}, {Wang}, {Wang}, {Wang}, {Wang}, {Wang}, {Wang}, {Wang}, {Wang}, {Wang}, {Wang}, {Wang}, {Xu}, {Xu}, {Yang}, {Yu}, {Yuan}, {Yuan}, {Zhai}, {Zhang}, {Zhang}, {Zhang}, {Zhao}, {Zhou}, {Zhou}, {Zhu}, \& {Zou}}]{LAMOST}
{Cui}, X.-Q., {Zhao}, Y.-H., {Chu}, Y.-Q., {et~al.} 2012, \href{http://dx.doi.org/10.1088/1674-4527/12/9/003}{\JournalTitle{Research in Astronomy and Astrophysics}, 12, 1197}

\bibitem[{{Cunningham} {et~al.}(2022){Cunningham}, {Sanderson}, {Johnston}, {Panithanpaisal}, {Ness}, {Wetzel}, {Loebman}, {Escala}, {Horta}, \& {Faucher-Gigu{\`e}re}}]{Cunningham2022}
{Cunningham}, E.~C., {Sanderson}, R.~E., {Johnston}, K.~V., {et~al.} 2022, \href{http://dx.doi.org/10.3847/1538-4357/ac78ea}{\JournalTitle{\apj}, 934, 172}

\bibitem[{{De Silva} {et~al.}(2015){De Silva}, {Freeman}, {Bland-Hawthorn}, {Martell}, {de Boer}, {Asplund}, {Keller}, {Sharma}, {Zucker}, {Zwitter}, {Anguiano}, {Bacigalupo}, {Bayliss}, {Beavis}, {Bergemann}, {Campbell}, {Cannon}, {Carollo}, {Casagrande}, {Casey}, {Da Costa}, {D'Orazi}, {Dotter}, {Duong}, {Heger}, {Ireland}, {Kafle}, {Kos}, {Lattanzio}, {Lewis}, {Lin}, {Lind}, {Munari}, {Nataf}, {O'Toole}, {Parker}, {Reid}, {Schlesinger}, {Sheinis}, {Simpson}, {Stello}, {Ting}, {Traven}, {Watson}, {Wittenmyer}, {Yong}, \& {{\v{Z}}erjal}}]{GALAH}
{De Silva}, G.~M., {Freeman}, K.~C., {Bland-Hawthorn}, J., {et~al.} 2015, \href{http://dx.doi.org/10.1093/mnras/stv327}{\JournalTitle{\mnras}, 449, 2604}

\bibitem[{{Emerick} {et~al.}(2020{\natexlab{a}}){Emerick}, {Bryan}, \& {Mac Low}}]{Emerick2020}
{Emerick}, A., {Bryan}, G.~L., \& {Mac Low}, M.-M. 2020{\natexlab{a}}, \href{http://dx.doi.org/10.3847/1538-4357/ab6efc}{\JournalTitle{\apj}, 890, 155}

\bibitem[{{Emerick} {et~al.}(2020{\natexlab{b}}){Emerick}, {Bryan}, \& {Mac Low}}]{Emerick+2020}
---. 2020{\natexlab{b}}, \href{http://dx.doi.org/10.48550/arXiv.2007.03702}{\JournalTitle{arXiv e-prints}, arXiv:2007.03702}

\bibitem[{{Emerick} {et~al.}(2018){Emerick}, {Bryan}, {Mac Low}, {C{\^o}t{\'e}}, {Johnston}, \& {O'Shea}}]{Emerick2018}
{Emerick}, A., {Bryan}, G.~L., {Mac Low}, M.-M., {et~al.} 2018, \href{http://dx.doi.org/10.3847/1538-4357/aaec7d}{\JournalTitle{\apj}, 869, 94}

\bibitem[{{Escala} {et~al.}(2018){Escala}, {Wetzel}, {Kirby}, {Hopkins}, {Ma}, {Wheeler}, {Kere{\v{s}}}, {Faucher-Gigu{\`e}re}, \& {Quataert}}]{Escala2018}
{Escala}, I., {Wetzel}, A., {Kirby}, E.~N., {et~al.} 2018, \href{http://dx.doi.org/10.1093/mnras/stx2858}{\JournalTitle{\mnras}, 474, 2194}

\bibitem[{{Fernandes} {et~al.}(2023){Fernandes}, {Mason}, {Horta}, {Schiavon}, {Hayes}, {Hasselquist}, {Feuillet}, {Beaton}, {J{\"o}nsson}, {Kisku}, {Lacerna}, {Lian}, {Minniti}, \& {Villanova}}]{Fernandes2023}
{Fernandes}, L., {Mason}, A.~C., {Horta}, D., {et~al.} 2023, \href{http://dx.doi.org/10.1093/mnras/stac3543}{\JournalTitle{\mnras}, 519, 3611}

\bibitem[{{Font} {et~al.}(2006){Font}, {Johnston}, {Bullock}, \& {Robertson}}]{Font2006}
{Font}, A.~S., {Johnston}, K.~V., {Bullock}, J.~S., \& {Robertson}, B.~E. 2006, \href{http://dx.doi.org/10.1086/498970}{\JournalTitle{\apj}, 638, 585}

\bibitem[{{Foreman-Mackey} {et~al.}(2013){Foreman-Mackey}, {Hogg}, {Lang}, \& {Goodman}}]{emcee}
{Foreman-Mackey}, D., {Hogg}, D.~W., {Lang}, D., \& {Goodman}, J. 2013, \href{http://dx.doi.org/10.1086/670067}{\JournalTitle{PASP}, 125, 306}

\bibitem[{{Frebel} \& {Bromm}(2012)}]{Frebel2012}
{Frebel}, A., \& {Bromm}, V. 2012, \href{http://dx.doi.org/10.1088/0004-637X/759/2/115}{\JournalTitle{\apj}, 759, 115}

\bibitem[{{Freeman} \& {Bland-Hawthorn}(2002)}]{Freeman2002}
{Freeman}, K., \& {Bland-Hawthorn}, J. 2002, \href{http://dx.doi.org/10.1146/annurev.astro.40.060401.093840}{\JournalTitle{\araa}, 40, 487}

\bibitem[{{Fu} {et~al.}(2023){Fu}, {Weisz}, {Starkenburg}, {Martin}, {Mercado}, {Savino}, {Boylan-Kolchin}, {C{\^o}t{\'e}}, {Dolphin}, {Longeard}, {Mateo}, {Samuel}, \& {Sandford}}]{Fu2023}
{Fu}, S.~W., {Weisz}, D.~R., {Starkenburg}, E., {et~al.} 2023, \href{http://dx.doi.org/10.48550/arXiv.2312.05981}{\JournalTitle{arXiv e-prints}, arXiv:2312.05981}

\bibitem[{{Gaia Collaboration} {et~al.}(2016){Gaia Collaboration}, {Prusti}, {de Bruijne}, {Brown}, {Vallenari}, {Babusiaux}, {Bailer-Jones}, {Bastian}, {Biermann}, {Evans}, {Eyer}, {Jansen}, {Jordi}, {Klioner}, {Lammers}, {Lindegren}, {Luri}, {Mignard}, {Milligan}, {Panem}, {Poinsignon}, {Pourbaix}, {Randich}, {Sarri}, {Sartoretti}, {Siddiqui}, {Soubiran}, {Valette}, {van Leeuwen}, {Walton}, {Aerts}, {Arenou}, {Cropper}, {Drimmel}, {H{\o}g}, {Katz}, {Lattanzi}, {O'Mullane}, {Grebel}, {Holland}, {Huc}, {Passot}, {Bramante}, {Cacciari}, {Casta{\~n}eda}, {Chaoul}, {Cheek}, {De Angeli}, {Fabricius}, {Guerra}, {Hern{\'a}ndez}, {Jean-Antoine-Piccolo}, {Masana}, {Messineo}, {Mowlavi}, {Nienartowicz}, {Ord{\'o}{\~n}ez-Blanco}, {Panuzzo}, {Portell}, {Richards}, {Riello}, {Seabroke}, {Tanga}, {Th{\'e}venin}, {Torra}, {Els}, {Gracia-Abril}, {Comoretto}, {Garcia-Reinaldos}, {Lock}, {Mercier}, {Altmann}, {Andrae}, {Astraatmadja}, {Bellas-Velidis}, {Benson}, {Berthier}, {Blomme}, {Busso}, {Carry}, {Cellino}, {Clementini},
  {Cowell}, {Creevey}, {Cuypers}, {Davidson}, {De Ridder}, {de Torres}, {Delchambre}, {Dell'Oro}, {Ducourant}, {Fr{\'e}mat}, {Garc{\'\i}a-Torres}, {Gosset}, {Halbwachs}, {Hambly}, {Harrison}, {Hauser}, {Hestroffer}, {Hodgkin}, {Huckle}, {Hutton}, {Jasniewicz}, {Jordan}, {Kontizas}, {Korn}, {Lanzafame}, {Manteiga}, {Moitinho}, {Muinonen}, {Osinde}, {Pancino}, {Pauwels}, {Petit}, {Recio-Blanco}, {Robin}, {Sarro}, {Siopis}, {Smith}, {Smith}, {Sozzetti}, {Thuillot}, {van Reeven}, {Viala}, {Abbas}, {Abreu Aramburu}, {Accart}, {Aguado}, {Allan}, {Allasia}, {Altavilla}, {{\'A}lvarez}, {Alves}, {Anderson}, {Andrei}, {Anglada Varela}, {Antiche}, {Antoja}, {Ant{\'o}n}, {Arcay}, {Atzei}, {Ayache}, {Bach}, {Baker}, {Balaguer-N{\'u}{\~n}ez}, {Barache}, {Barata}, {Barbier}, {Barblan}, {Baroni}, {Barrado y Navascu{\'e}s}, {Barros}, {Barstow}, {Becciani}, {Bellazzini}, {Bellei}, {Bello Garc{\'\i}a}, {Belokurov}, {Bendjoya}, {Berihuete}, {Bianchi}, {Bienaym{\'e}}, {Billebaud}, {Blagorodnova}, {Blanco-Cuaresma}, {Boch},
  {Bombrun}, {Borrachero}, {Bouquillon}, {Bourda}, {Bouy}, {Bragaglia}, {Breddels}, {Brouillet}, {Br{\"u}semeister}, {Bucciarelli}, {Budnik}, {Burgess}, {Burgon}, {Burlacu}, {Busonero}, {Buzzi}, {Caffau}, {Cambras}, {Campbell}, {Cancelliere}, {Cantat-Gaudin}, {Carlucci}, {Carrasco}, {Castellani}, {Charlot}, {Charnas}, {Charvet}, {Chassat}, {Chiavassa}, {Clotet}, {Cocozza}, {Collins}, {Collins}, {Costigan}, {Crifo}, {Cross}, {Crosta}, {Crowley}, {Dafonte}, {Damerdji}, {Dapergolas}, {David}, {David}, {De Cat}, {de Felice}, {de Laverny}, {De Luise}, {De March}, {de Martino}, {de Souza}, {Debosscher}, {del Pozo}, {Delbo}, {Delgado}, {Delgado}, {di Marco}, {Di Matteo}, {Diakite}, {Distefano}, {Dolding}, {Dos Anjos}, {Drazinos}, {Dur{\'a}n}, {Dzigan}, {Ecale}, {Edvardsson}, {Enke}, {Erdmann}, {Escolar}, {Espina}, {Evans}, {Eynard Bontemps}, {Fabre}, {Fabrizio}, {Faigler}, {Falc{\~a}o}, {Farr{\`a}s Casas}, {Faye}, {Federici}, {Fedorets}, {Fern{\'a}ndez-Hern{\'a}ndez}, {Fernique}, {Fienga}, {Figueras}, {Filippi},
  {Findeisen}, {Fonti}, {Fouesneau}, {Fraile}, {Fraser}, {Fuchs}, {Furnell}, {Gai}, {Galleti}, {Galluccio}, {Garabato}, {Garc{\'\i}a-Sedano}, {Gar{\'e}}, {Garofalo}, {Garralda}, {Gavras}, {Gerssen}, {Geyer}, {Gilmore}, {Girona}, {Giuffrida}, {Gomes}, {Gonz{\'a}lez-Marcos}, {Gonz{\'a}lez-N{\'u}{\~n}ez}, {Gonz{\'a}lez-Vidal}, {Granvik}, {Guerrier}, {Guillout}, {Guiraud}, {G{\'u}rpide}, {Guti{\'e}rrez-S{\'a}nchez}, {Guy}, {Haigron}, {Hatzidimitriou}, {Haywood}, {Heiter}, {Helmi}, {Hobbs}, {Hofmann}, {Holl}, {Holland}, {Hunt}, {Hypki}, {Icardi}, {Irwin}, {Jevardat de Fombelle}, {Jofr{\'e}}, {Jonker}, {Jorissen}, {Julbe}, {Karampelas}, {Kochoska}, {Kohley}, {Kolenberg}, {Kontizas}, {Koposov}, {Kordopatis}, {Koubsky}, {Kowalczyk}, {Krone-Martins}, {Kudryashova}, {Kull}, {Bachchan}, {Lacoste-Seris}, {Lanza}, {Lavigne}, {Le Poncin-Lafitte}, {Lebreton}, {Lebzelter}, {Leccia}, {Leclerc}, {Lecoeur-Taibi}, {Lemaitre}, {Lenhardt}, {Leroux}, {Liao}, {Licata}, {Lindstr{\o}m}, {Lister}, {Livanou}, {Lobel}, {L{\"o}ffler},
  {L{\'o}pez}, {Lopez-Lozano}, {Lorenz}, {Loureiro}, {MacDonald}, {Magalh{\~a}es Fernandes}, {Managau}, {Mann}, {Mantelet}, {Marchal}, {Marchant}, {Marconi}, {Marie}, {Marinoni}, {Marrese}, {Marschalk{\'o}}, {Marshall}, {Mart{\'\i}n-Fleitas}, {Martino}, {Mary}, {Matijevi{\v{c}}}, {Mazeh}, {McMillan}, {Messina}, {Mestre}, {Michalik}, {Millar}, {Miranda}, {Molina}, {Molinaro}, {Molinaro}, {Moln{\'a}r}, {Moniez}, {Montegriffo}, {Monteiro}, {Mor}, {Mora}, {Morbidelli}, {Morel}, {Morgenthaler}, {Morley}, {Morris}, {Mulone}, {Muraveva}, {Musella}, {Narbonne}, {Nelemans}, {Nicastro}, {Noval}, {Ord{\'e}novic}, {Ordieres-Mer{\'e}}, {Osborne}, {Pagani}, {Pagano}, {Pailler}, {Palacin}, {Palaversa}, {Parsons}, {Paulsen}, {Pecoraro}, {Pedrosa}, {Pentik{\"a}inen}, {Pereira}, {Pichon}, {Piersimoni}, {Pineau}, {Plachy}, {Plum}, {Poujoulet}, {Pr{\v{s}}a}, {Pulone}, {Ragaini}, {Rago}, {Rambaux}, {Ramos-Lerate}, {Ranalli}, {Rauw}, {Read}, {Regibo}, {Renk}, {Reyl{\'e}}, {Ribeiro}, {Rimoldini}, {Ripepi}, {Riva}, {Rixon},
  {Roelens}, {Romero-G{\'o}mez}, {Rowell}, {Royer}, {Rudolph}, {Ruiz-Dern}, {Sadowski}, {Sagrist{\`a} Sell{\'e}s}, {Sahlmann}, {Salgado}, {Salguero}, {Sarasso}, {Savietto}, {Schnorhk}, {Schultheis}, {Sciacca}, {Segol}, {Segovia}, {Segransan}, {Serpell}, {Shih}, {Smareglia}, {Smart}, {Smith}, {Solano}, {Solitro}, {Sordo}, {Soria Nieto}, {Souchay}, {Spagna}, {Spoto}, {Stampa}, {Steele}, {Steidelm{\"u}ller}, {Stephenson}, {Stoev}, {Suess}, {S{\"u}veges}, {Surdej}, {Szabados}, {Szegedi-Elek}, {Tapiador}, {Taris}, {Tauran}, {Taylor}, {Teixeira}, {Terrett}, {Tingley}, {Trager}, {Turon}, {Ulla}, {Utrilla}, {Valentini}, {van Elteren}, {Van Hemelryck}, {van Leeuwen}, {Varadi}, {Vecchiato}, {Veljanoski}, {Via}, {Vicente}, {Vogt}, {Voss}, {Votruba}, {Voutsinas}, {Walmsley}, {Weiler}, {Weingrill}, {Werner}, {Wevers}, {Whitehead}, {Wyrzykowski}, {Yoldas}, {{\v{Z}}erjal}, {Zucker}, {Zurbach}, {Zwitter}, {Alecu}, {Allen}, {Allende Prieto}, {Amorim}, {Anglada-Escud{\'e}}, {Arsenijevic}, {Azaz}, {Balm}, {Beck}, {Bernstein},
  {Bigot}, {Bijaoui}, {Blasco}, {Bonfigli}, {Bono}, {Boudreault}, {Bressan}, {Brown}, {Brunet}, {Bunclark}, {Buonanno}, {Butkevich}, {Carret}, {Carrion}, {Chemin}, {Ch{\'e}reau}, {Corcione}, {Darmigny}, {de Boer}, {de Teodoro}, {de Zeeuw}, {Delle Luche}, {Domingues}, {Dubath}, {Fodor}, {Fr{\'e}zouls}, {Fries}, {Fustes}, {Fyfe}, {Gallardo}, {Gallegos}, {Gardiol}, {Gebran}, {Gomboc}, {G{\'o}mez}, {Grux}, {Gueguen}, {Heyrovsky}, {Hoar}, {Iannicola}, {Isasi Parache}, {Janotto}, {Joliet}, {Jonckheere}, {Keil}, {Kim}, {Klagyivik}, {Klar}, {Knude}, {Kochukhov}, {Kolka}, {Kos}, {Kutka}, {Lainey}, {LeBouquin}, {Liu}, {Loreggia}, {Makarov}, {Marseille}, {Martayan}, {Martinez-Rubi}, {Massart}, {Meynadier}, {Mignot}, {Munari}, {Nguyen}, {Nordlander}, {Ocvirk}, {O'Flaherty}, {Olias Sanz}, {Ortiz}, {Osorio}, {Oszkiewicz}, {Ouzounis}, {Palmer}, {Park}, {Pasquato}, {Peltzer}, {Peralta}, {P{\'e}turaud}, {Pieniluoma}, {Pigozzi}, {Poels}, {Prat}, {Prod'homme}, {Raison}, {Rebordao}, {Risquez}, {Rocca-Volmerange}, {Rosen},
  {Ruiz-Fuertes}, {Russo}, {Sembay}, {Serraller Vizcaino}, {Short}, {Siebert}, {Silva}, {Sinachopoulos}, {Slezak}, {Soffel}, {Sosnowska}, {Strai{\v{z}}ys}, {ter Linden}, {Terrell}, {Theil}, {Tiede}, {Troisi}, {Tsalmantza}, {Tur}, {Vaccari}, {Vachier}, {Valles}, {Van Hamme}, {Veltz}, {Virtanen}, {Wallut}, {Wichmann}, {Wilkinson}, {Ziaeepour}, \& {Zschocke}}]{GaiaMission}
{Gaia Collaboration}, {Prusti}, T., {de Bruijne}, J.~H.~J., {et~al.} 2016, \href{http://dx.doi.org/10.1051/0004-6361/201629272}{\JournalTitle{\aap}, 595, A1}

\bibitem[{{Gaia Collaboration} {et~al.}(2023){Gaia Collaboration}, {Vallenari}, {Brown}, {Prusti}, {de Bruijne}, {Arenou}, {Babusiaux}, {Biermann}, {Creevey}, {Ducourant}, {Evans}, {Eyer}, {Guerra}, {Hutton}, {Jordi}, {Klioner}, {Lammers}, {Lindegren}, {Luri}, {Mignard}, {Panem}, {Pourbaix}, {Randich}, {Sartoretti}, {Soubiran}, {Tanga}, {Walton}, {Bailer-Jones}, {Bastian}, {Drimmel}, {Jansen}, {Katz}, {Lattanzi}, {van Leeuwen}, {Bakker}, {Cacciari}, {Casta{\~n}eda}, {De Angeli}, {Fabricius}, {Fouesneau}, {Fr{\'e}mat}, {Galluccio}, {Guerrier}, {Heiter}, {Masana}, {Messineo}, {Mowlavi}, {Nicolas}, {Nienartowicz}, {Pailler}, {Panuzzo}, {Riclet}, {Roux}, {Seabroke}, {Sordo}, {Th{\'e}venin}, {Gracia-Abril}, {Portell}, {Teyssier}, {Altmann}, {Andrae}, {Audard}, {Bellas-Velidis}, {Benson}, {Berthier}, {Blomme}, {Burgess}, {Busonero}, {Busso}, {C{\'a}novas}, {Carry}, {Cellino}, {Cheek}, {Clementini}, {Damerdji}, {Davidson}, {de Teodoro}, {Nu{\~n}ez Campos}, {Delchambre}, {Dell'Oro}, {Esquej},
  {Fern{\'a}ndez-Hern{\'a}ndez}, {Fraile}, {Garabato}, {Garc{\'\i}a-Lario}, {Gosset}, {Haigron}, {Halbwachs}, {Hambly}, {Harrison}, {Hern{\'a}ndez}, {Hestroffer}, {Hodgkin}, {Holl}, {Jan{\ss}en}, {Jevardat de Fombelle}, {Jordan}, {Krone-Martins}, {Lanzafame}, {L{\"o}ffler}, {Marchal}, {Marrese}, {Moitinho}, {Muinonen}, {Osborne}, {Pancino}, {Pauwels}, {Recio-Blanco}, {Reyl{\'e}}, {Riello}, {Rimoldini}, {Roegiers}, {Rybizki}, {Sarro}, {Siopis}, {Smith}, {Sozzetti}, {Utrilla}, {van Leeuwen}, {Abbas}, {{\'A}brah{\'a}m}, {Abreu Aramburu}, {Aerts}, {Aguado}, {Ajaj}, {Aldea-Montero}, {Altavilla}, {{\'A}lvarez}, {Alves}, {Anders}, {Anderson}, {Anglada Varela}, {Antoja}, {Baines}, {Baker}, {Balaguer-N{\'u}{\~n}ez}, {Balbinot}, {Balog}, {Barache}, {Barbato}, {Barros}, {Barstow}, {Bartolom{\'e}}, {Bassilana}, {Bauchet}, {Becciani}, {Bellazzini}, {Berihuete}, {Bernet}, {Bertone}, {Bianchi}, {Binnenfeld}, {Blanco-Cuaresma}, {Blazere}, {Boch}, {Bombrun}, {Bossini}, {Bouquillon}, {Bragaglia}, {Bramante}, {Breedt},
  {Bressan}, {Brouillet}, {Brugaletta}, {Bucciarelli}, {Burlacu}, {Butkevich}, {Buzzi}, {Caffau}, {Cancelliere}, {Cantat-Gaudin}, {Carballo}, {Carlucci}, {Carnerero}, {Carrasco}, {Casamiquela}, {Castellani}, {Castro-Ginard}, {Chaoul}, {Charlot}, {Chemin}, {Chiaramida}, {Chiavassa}, {Chornay}, {Comoretto}, {Contursi}, {Cooper}, {Cornez}, {Cowell}, {Crifo}, {Cropper}, {Crosta}, {Crowley}, {Dafonte}, {Dapergolas}, {David}, {David}, {de Laverny}, {De Luise}, {De March}, {De Ridder}, {de Souza}, {de Torres}, {del Peloso}, {del Pozo}, {Delbo}, {Delgado}, {Delisle}, {Demouchy}, {Dharmawardena}, {Di Matteo}, {Diakite}, {Diener}, {Distefano}, {Dolding}, {Edvardsson}, {Enke}, {Fabre}, {Fabrizio}, {Faigler}, {Fedorets}, {Fernique}, {Fienga}, {Figueras}, {Fournier}, {Fouron}, {Fragkoudi}, {Gai}, {Garcia-Gutierrez}, {Garcia-Reinaldos}, {Garc{\'\i}a-Torres}, {Garofalo}, {Gavel}, {Gavras}, {Gerlach}, {Geyer}, {Giacobbe}, {Gilmore}, {Girona}, {Giuffrida}, {Gomel}, {Gomez}, {Gonz{\'a}lez-N{\'u}{\~n}ez},
  {Gonz{\'a}lez-Santamar{\'\i}a}, {Gonz{\'a}lez-Vidal}, {Granvik}, {Guillout}, {Guiraud}, {Guti{\'e}rrez-S{\'a}nchez}, {Guy}, {Hatzidimitriou}, {Hauser}, {Haywood}, {Helmer}, {Helmi}, {Sarmiento}, {Hidalgo}, {Hilger}, {H{\l}adczuk}, {Hobbs}, {Holland}, {Huckle}, {Jardine}, {Jasniewicz}, {Jean-Antoine Piccolo}, {Jim{\'e}nez-Arranz}, {Jorissen}, {Juaristi Campillo}, {Julbe}, {Karbevska}, {Kervella}, {Khanna}, {Kontizas}, {Kordopatis}, {Korn}, {K{\'o}sp{\'a}l}, {Kostrzewa-Rutkowska}, {Kruszy{\'n}ska}, {Kun}, {Laizeau}, {Lambert}, {Lanza}, {Lasne}, {Le Campion}, {Lebreton}, {Lebzelter}, {Leccia}, {Leclerc}, {Lecoeur-Taibi}, {Liao}, {Licata}, {Lindstr{\o}m}, {Lister}, {Livanou}, {Lobel}, {Lorca}, {Loup}, {Madrero Pardo}, {Magdaleno Romeo}, {Managau}, {Mann}, {Manteiga}, {Marchant}, {Marconi}, {Marcos}, {Marcos Santos}, {Mar{\'\i}n Pina}, {Marinoni}, {Marocco}, {Marshall}, {Martin Polo}, {Mart{\'\i}n-Fleitas}, {Marton}, {Mary}, {Masip}, {Massari}, {Mastrobuono-Battisti}, {Mazeh}, {McMillan}, {Messina}, {Michalik},
  {Millar}, {Mints}, {Molina}, {Molinaro}, {Moln{\'a}r}, {Monari}, {Mongui{\'o}}, {Montegriffo}, {Montero}, {Mor}, {Mora}, {Morbidelli}, {Morel}, {Morris}, {Muraveva}, {Murphy}, {Musella}, {Nagy}, {Noval}, {Oca{\~n}a}, {Ogden}, {Ordenovic}, {Osinde}, {Pagani}, {Pagano}, {Palaversa}, {Palicio}, {Pallas-Quintela}, {Panahi}, {Payne-Wardenaar}, {Pe{\~n}alosa Esteller}, {Penttil{\"a}}, {Pichon}, {Piersimoni}, {Pineau}, {Plachy}, {Plum}, {Poggio}, {Pr{\v{s}}a}, {Pulone}, {Racero}, {Ragaini}, {Rainer}, {Raiteri}, {Rambaux}, {Ramos}, {Ramos-Lerate}, {Re Fiorentin}, {Regibo}, {Richards}, {Rios Diaz}, {Ripepi}, {Riva}, {Rix}, {Rixon}, {Robichon}, {Robin}, {Robin}, {Roelens}, {Rogues}, {Rohrbasser}, {Romero-G{\'o}mez}, {Rowell}, {Royer}, {Ruz Mieres}, {Rybicki}, {Sadowski}, {S{\'a}ez N{\'u}{\~n}ez}, {Sagrist{\`a} Sell{\'e}s}, {Sahlmann}, {Salguero}, {Samaras}, {Sanchez Gimenez}, {Sanna}, {Santove{\~n}a}, {Sarasso}, {Schultheis}, {Sciacca}, {Segol}, {Segovia}, {S{\'e}gransan}, {Semeux}, {Shahaf}, {Siddiqui}, {Siebert},
  {Siltala}, {Silvelo}, {Slezak}, {Slezak}, {Smart}, {Snaith}, {Solano}, {Solitro}, {Souami}, {Souchay}, {Spagna}, {Spina}, {Spoto}, {Steele}, {Steidelm{\"u}ller}, {Stephenson}, {S{\"u}veges}, {Surdej}, {Szabados}, {Szegedi-Elek}, {Taris}, {Taylor}, {Teixeira}, {Tolomei}, {Tonello}, {Torra}, {Torra}, {Torralba Elipe}, {Trabucchi}, {Tsounis}, {Turon}, {Ulla}, {Unger}, {Vaillant}, {van Dillen}, {van Reeven}, {Vanel}, {Vecchiato}, {Viala}, {Vicente}, {Voutsinas}, {Weiler}, {Wevers}, {Wyrzykowski}, {Yoldas}, {Yvard}, {Zhao}, {Zorec}, {Zucker}, \& {Zwitter}}]{GaiaDR3}
{Gaia Collaboration}, {Vallenari}, A., {Brown}, A.~G.~A., {et~al.} 2023, \href{http://dx.doi.org/10.1051/0004-6361/202243940}{\JournalTitle{\aap}, 674, A1}

\bibitem[{{Garc{\'\i}a P{\'e}rez} {et~al.}(2016){Garc{\'\i}a P{\'e}rez}, {Allende Prieto}, {Holtzman}, {Shetrone}, {M{\'e}sz{\'a}ros}, {Bizyaev}, {Carrera}, {Cunha}, {Garc{\'\i}a-Hern{\'a}ndez}, {Johnson}, {Majewski}, {Nidever}, {Schiavon}, {Shane}, {Smith}, {Sobeck}, {Troup}, {Zamora}, {Weinberg}, {Bovy}, {Eisenstein}, {Feuillet}, {Frinchaboy}, {Hayden}, {Hearty}, {Nguyen}, {O'Connell}, {Pinsonneault}, {Wilson}, \& {Zasowski}}]{aspcap}
{Garc{\'\i}a P{\'e}rez}, A.~E., {Allende Prieto}, C., {Holtzman}, J.~A., {et~al.} 2016, \href{http://dx.doi.org/10.3847/0004-6256/151/6/144}{\JournalTitle{\aj}, 151, 144}

\bibitem[{{Geha} {et~al.}(2013){Geha}, {Brown}, {Tumlinson}, {Kalirai}, {Simon}, {Kirby}, {VandenBerg}, {Mu{\~n}oz}, {Avila}, {Guhathakurta}, \& {Ferguson}}]{Geha2013}
{Geha}, M., {Brown}, T.~M., {Tumlinson}, J., {et~al.} 2013, \href{http://dx.doi.org/10.1088/0004-637X/771/1/29}{\JournalTitle{\apj}, 771, 29}

\bibitem[{{Genina} {et~al.}(2019){Genina}, {Frenk}, {Ben{\'\i}tez-Llambay}, {Cole}, {Navarro}, {Oman}, \& {Fattahi}}]{Genina2019}
{Genina}, A., {Frenk}, C.~S., {Ben{\'\i}tez-Llambay}, A., {et~al.} 2019, \href{http://dx.doi.org/10.1093/mnras/stz1852}{\JournalTitle{\mnras}, 488, 2312}

\bibitem[{{Goriely} {et~al.}(2011){Goriely}, {Bauswein}, \& {Janka}}]{Goriely2011}
{Goriely}, S., {Bauswein}, A., \& {Janka}, H.-T. 2011, \href{http://dx.doi.org/10.1088/2041-8205/738/2/L32}{\JournalTitle{\apjl}, 738, L32}

\bibitem[{{Gratton}(2020)}]{Gratton2020}
{Gratton}, R. 2020, \href{http://dx.doi.org/10.1017/S1743921319007877}{in Star Clusters: From the Milky Way to the Early Universe, ed. A.~{Bragaglia}, M.~{Davies}, A.~{Sills}, \& E.~{Vesperini}, Vol. 351}, 241

\bibitem[{{Griffith} {et~al.}(2021){Griffith}, {Weinberg}, {Johnson}, {Beaton}, {Garc{\'\i}a-Hern{\'a}ndez}, {Hasselquist}, {Holtzman}, {Johnson}, {J{\"o}nsson}, {Lane}, {Nataf}, \& {Roman-Lopes}}]{Griffith2021a}
{Griffith}, E., {Weinberg}, D.~H., {Johnson}, J.~A., {et~al.} 2021, \href{http://dx.doi.org/10.3847/1538-4357/abd6be}{\JournalTitle{\apj}, 909, 77}

\bibitem[{{Griffith} {et~al.}(2023){Griffith}, {Johnson}, {Weinberg}, {Ilyin}, {Johnson}, {Rodriguez-Martinez}, \& {Strassmeier}}]{Griffith23}
{Griffith}, E.~J., {Johnson}, J.~A., {Weinberg}, D.~H., {et~al.} 2023, \href{http://dx.doi.org/10.3847/1538-4357/aca659}{\JournalTitle{\apj}, 944, 47}

\bibitem[{{Gutcke} {et~al.}(2021){Gutcke}, {Pakmor}, {Naab}, \& {Springel}}]{Gutcke+2021}
{Gutcke}, T.~A., {Pakmor}, R., {Naab}, T., \& {Springel}, V. 2021, \href{http://dx.doi.org/10.1093/mnras/staa3875}{\JournalTitle{\mnras}, 501, 5597}

\bibitem[{{Harris}(1996)}]{Harris1996}
{Harris}, W.~E. 1996, \href{http://dx.doi.org/10.1086/118116}{\JournalTitle{\aj}, 112, 1487}

\bibitem[{{Hartwig} {et~al.}(2019){Hartwig}, {Ishigaki}, {Klessen}, \& {Yoshida}}]{Hartwig2019}
{Hartwig}, T., {Ishigaki}, M.~N., {Klessen}, R.~S., \& {Yoshida}, N. 2019, \href{http://dx.doi.org/10.1093/mnras/sty2783}{\JournalTitle{\mnras}, 482, 1204}

\bibitem[{{Hartwig} {et~al.}(2018){Hartwig}, {Yoshida}, {Magg}, {Frebel}, {Glover}, {G{\'o}mez}, {Griffen}, {Ishigaki}, {Ji}, {Klessen}, {O'Shea}, \& {Tominaga}}]{Hartwig2018}
{Hartwig}, T., {Yoshida}, N., {Magg}, M., {et~al.} 2018, \href{http://dx.doi.org/10.1093/mnras/sty1176}{\JournalTitle{\mnras}, 478, 1795}

\bibitem[{{Hasselquist} {et~al.}(2021){Hasselquist}, {Hayes}, {Lian}, {Weinberg}, {Zasowski}, {Horta}, {Beaton}, {Feuillet}, {Garro}, {Gallart}, {Smith}, {Holtzman}, {Minniti}, {Lacerna}, {Shetrone}, {J{\"o}nsson}, {Cioni}, {Fillingham}, {Cunha}, {O'Connell}, {Fern{\'a}ndez-Trincado}, {Mu{\~n}oz}, {Schiavon}, {Almeida}, {Anguiano}, {Beers}, {Bizyaev}, {Brownstein}, {Cohen}, {Frinchaboy}, {Garc{\'\i}a-Hern{\'a}ndez}, {Geisler}, {Lane}, {Majewski}, {Nidever}, {Nitschelm}, {Povick}, {Price-Whelan}, {Roman-Lopes}, {Rosado}, {Sobeck}, {Stringfellow}, {Valenzuela}, {Villanova}, \& {Vincenzo}}]{Hasselquist2021}
{Hasselquist}, S., {Hayes}, C.~R., {Lian}, J., {et~al.} 2021, \href{http://dx.doi.org/10.3847/1538-4357/ac25f9}{\JournalTitle{\apj}, 923, 172}

\bibitem[{{Hayden} {et~al.}(2015){Hayden}, {Bovy}, {Holtzman}, {Nidever}, {Bird}, {Weinberg}, {Andrews}, {Majewski}, {Allende Prieto}, {Anders}, {Beers}, {Bizyaev}, {Chiappini}, {Cunha}, {Frinchaboy}, {Garc{\'\i}a-Her{\'n}andez}, {Garc{\'\i}a P{\'e}rez}, {Girardi}, {Harding}, {Hearty}, {Johnson}, {M{\'e}sz{\'a}ros}, {Minchev}, {O'Connell}, {Pan}, {Robin}, {Schiavon}, {Schneider}, {Schultheis}, {Shetrone}, {Skrutskie}, {Steinmetz}, {Smith}, {Wilson}, {Zamora}, \& {Zasowski}}]{Hayden2015}
{Hayden}, M.~R., {Bovy}, J., {Holtzman}, J.~A., {et~al.} 2015, \href{http://dx.doi.org/10.1088/0004-637X/808/2/132}{\JournalTitle{\apj}, 808, 132}

\bibitem[{{Herwig}(2005)}]{Herwig2005}
{Herwig}, F. 2005, \href{http://dx.doi.org/10.1146/annurev.astro.43.072103.150600}{\JournalTitle{\araa}, 43, 435}

\bibitem[{{Hill} {et~al.}(2019){Hill}, {Sk{\'u}lad{\'o}ttir}, {Tolstoy}, {Venn}, {Shetrone}, {Jablonka}, {Primas}, {Battaglia}, {de Boer}, {Fran{\c{c}}ois}, {Helmi}, {Kaufer}, {Letarte}, {Starkenburg}, \& {Spite}}]{Hill2019}
{Hill}, V., {Sk{\'u}lad{\'o}ttir}, {\'A}., {Tolstoy}, E., {et~al.} 2019, \href{http://dx.doi.org/10.1051/0004-6361/201833950}{\JournalTitle{\aap}, 626, A15}

\bibitem[{{Ho} {et~al.}(2017){Ho}, {Ness}, {Hogg}, {Rix}, {Liu}, {Yang}, {Zhang}, {Hou}, \& {Wang}}]{Ho2017}
{Ho}, A. Y.~Q., {Ness}, M.~K., {Hogg}, D.~W., {et~al.} 2017, \href{http://dx.doi.org/10.3847/1538-4357/836/1/5}{\JournalTitle{\apj}, 836, 5}

\bibitem[{{Holtzman} {et~al.}(2015){Holtzman}, {Shetrone}, {Johnson}, {Allende Prieto}, {Anders}, {Andrews}, {Beers}, {Bizyaev}, {Blanton}, {Bovy}, {Carrera}, {Chojnowski}, {Cunha}, {Eisenstein}, {Feuillet}, {Frinchaboy}, {Galbraith-Frew}, {Garc{\'\i}a P{\'e}rez}, {Garc{\'\i}a-Hern{\'a}ndez}, {Hasselquist}, {Hayden}, {Hearty}, {Ivans}, {Majewski}, {Martell}, {Meszaros}, {Muna}, {Nidever}, {Nguyen}, {O'Connell}, {Pan}, {Pinsonneault}, {Robin}, {Schiavon}, {Shane}, {Sobeck}, {Smith}, {Troup}, {Weinberg}, {Wilson}, {Wood-Vasey}, {Zamora}, \& {Zasowski}}]{Holtzman2015}
{Holtzman}, J.~A., {Shetrone}, M., {Johnson}, J.~A., {et~al.} 2015, \href{http://dx.doi.org/10.1088/0004-6256/150/5/148}{\JournalTitle{\aj}, 150, 148}

\bibitem[{{Holtzman} {et~al.}(2018){Holtzman}, {Hasselquist}, {Shetrone}, {Cunha}, {Allende Prieto}, {Anguiano}, {Bizyaev}, {Bovy}, {Casey}, {Edvardsson}, {Johnson}, {J{\"o}nsson}, {Meszaros}, {Smith}, {Sobeck}, {Zamora}, {Chojnowski}, {Fernandez-Trincado}, {Garcia-Hernandez}, {Majewski}, {Pinsonneault}, {Souto}, {Stringfellow}, {Tayar}, {Troup}, \& {Zasowski}}]{APOGEEDR13and14}
{Holtzman}, J.~A., {Hasselquist}, S., {Shetrone}, M., {et~al.} 2018, \href{http://dx.doi.org/10.3847/1538-3881/aad4f9}{\JournalTitle{\aj}, 156, 125}

\bibitem[{{Horta} {et~al.}(2023{\natexlab{a}}){Horta}, {Schiavon}, {Mackereth}, {Weinberg}, {Hasselquist}, {Feuillet}, {O'Connell}, {Anguiano}, {Allende-Prieto}, {Beaton}, {Bizyaev}, {Cunha}, {Geisler}, {Garc{\'\i}a-Hern{\'a}ndez}, {Holtzman}, {J{\"o}nsson}, {Lane}, {Majewski}, {M{\'e}sz{\'a}ros}, {Minniti}, {Nitschelm}, {Shetrone}, {Smith}, \& {Zasowski}}]{Horta2023}
{Horta}, D., {Schiavon}, R.~P., {Mackereth}, J.~T., {et~al.} 2023{\natexlab{a}}, \href{http://dx.doi.org/10.1093/mnras/stac3179}{\JournalTitle{\mnras}, 520, 5671}

\bibitem[{{Horta} {et~al.}(2023{\natexlab{b}}){Horta}, {Cunningham}, {Sanderson}, {Johnston}, {Panithanpaisal}, {Arora}, {Necib}, {Wetzel}, {Bailin}, \& {Faucher-Gigu{\`e}re}}]{HortaCunningham2023}
{Horta}, D., {Cunningham}, E.~C., {Sanderson}, R.~E., {et~al.} 2023{\natexlab{b}}, \href{http://dx.doi.org/10.3847/1538-4357/acae87}{\JournalTitle{\apj}, 943, 158}

\bibitem[{{Ji} {et~al.}(2019){Ji}, {Beaton}, {Chakrabarti}, {Duggan}, {Frebel}, {Geha}, {Hosek}, {Kirby}, {Li}, {Roederer}, \& {Simon}}]{Ji2019}
{Ji}, A., {Beaton}, R., {Chakrabarti}, S., {et~al.} 2019, \href{http://dx.doi.org/10.48550/arXiv.1903.09275}{\JournalTitle{\baas}, 51, 166}

\bibitem[{{Ji} {et~al.}(2015){Ji}, {Frebel}, \& {Bromm}}]{Ji2015}
{Ji}, A.~P., {Frebel}, A., \& {Bromm}, V. 2015, \href{http://dx.doi.org/10.1093/mnras/stv2052}{\JournalTitle{\mnras}, 454, 659}

\bibitem[{{Ji} {et~al.}(2020){Ji}, {Li}, {Simon}, {Marshall}, {Vivas}, {Pace}, {Bechtol}, {Drlica-Wagner}, {Koposov}, {Hansen}, {Allam}, {Gruendl}, {Johnson}, {McNanna}, {No{\"e}l}, {Tucker}, \& {Walker}}]{Ji2020}
{Ji}, A.~P., {Li}, T.~S., {Simon}, J.~D., {et~al.} 2020, \href{http://dx.doi.org/10.3847/1538-4357/ab6213}{\JournalTitle{\apj}, 889, 27}

\bibitem[{{Ji} {et~al.}(2022){Ji}, {Simon}, {Roederer}, {Magg}, {Frebel}, {Johnson}, {Klessen}, {Magg}, {Cescutti}, {Mateo}, {Bergemann}, \& {Bailey}}]{Ji2022}
{Ji}, A.~P., {Simon}, J.~D., {Roederer}, I.~U., {et~al.} 2022, \JournalTitle{arXiv e-prints}, arXiv:2207.03499

\bibitem[{{Ji} {et~al.}(2023){Ji}, {Simon}, {Roederer}, {Magg}, {Frebel}, {Johnson}, {Klessen}, {Magg}, {Cescutti}, {Mateo}, {Bergemann}, \& {Bailey}}]{Ji2023}
---. 2023, \href{http://dx.doi.org/10.3847/1538-3881/acad84}{\JournalTitle{\aj}, 165, 100}

\bibitem[{{Johnston} {et~al.}(2008){Johnston}, {Bullock}, {Sharma}, {Font}, {Robertson}, \& {Leitner}}]{Johnston2008}
{Johnston}, K.~V., {Bullock}, J.~S., {Sharma}, S., {et~al.} 2008, \href{http://dx.doi.org/10.1086/592228}{\JournalTitle{\apj}, 689, 936}

\bibitem[{{J{\"o}nsson} {et~al.}(2020){J{\"o}nsson}, {Holtzman}, {Allende Prieto}, {Cunha}, {Garc{\'\i}a-Hern{\'a}ndez}, {Hasselquist}, {Masseron}, {Osorio}, {Shetrone}, {Smith}, {Stringfellow}, {Bizyaev}, {Edvardsson}, {Majewski}, {M{\'e}sz{\'a}ros}, {Souto}, {Zamora}, {Beaton}, {Bovy}, {Donor}, {Pinsonneault}, {Poovelil}, \& {Sobeck}}]{Jonsson2020}
{J{\"o}nsson}, H., {Holtzman}, J.~A., {Allende Prieto}, C., {et~al.} 2020, \href{http://dx.doi.org/10.3847/1538-3881/aba592}{\JournalTitle{\aj}, 160, 120}

\bibitem[{{Karakas}(2010)}]{Karakas2010}
{Karakas}, A.~I. 2010, \href{http://dx.doi.org/10.1111/j.1365-2966.2009.16198.x}{\JournalTitle{\mnras}, 403, 1413}

\bibitem[{{Karakas} \& {Lattanzio}(2014)}]{Karakas2014}
{Karakas}, A.~I., \& {Lattanzio}, J.~C. 2014, \href{http://dx.doi.org/10.1017/pasa.2014.21}{\JournalTitle{\pasa}, 31, e030}

\bibitem[{{Kirby} {et~al.}(2013){Kirby}, {Cohen}, {Guhathakurta}, {Cheng}, {Bullock}, \& {Gallazzi}}]{Kirby2013}
{Kirby}, E.~N., {Cohen}, J.~G., {Guhathakurta}, P., {et~al.} 2013, \href{http://dx.doi.org/10.1088/0004-637X/779/2/102}{\JournalTitle{\apj}, 779, 102}

\bibitem[{{Kirby} {et~al.}(2008){Kirby}, {Simon}, {Geha}, {Guhathakurta}, \& {Frebel}}]{Kirby2008}
{Kirby}, E.~N., {Simon}, J.~D., {Geha}, M., {Guhathakurta}, P., \& {Frebel}, A. 2008, \href{http://dx.doi.org/10.1086/592432}{\JournalTitle{\apjl}, 685, L43}

\bibitem[{{Kirby} {et~al.}(2018){Kirby}, {Xie}, {Guo}, {Kovalev}, \& {Bergemann}}]{Kirby2018_data}
{Kirby}, E.~N., {Xie}, J.~L., {Guo}, R., {Kovalev}, M., \& {Bergemann}, M. 2018, \JournalTitle{VizieR Online Data Catalog}, J/ApJS/237/18

\bibitem[{{Kobayashi} {et~al.}(2020){Kobayashi}, {Karakas}, \& {Lugaro}}]{Kobayashi2020}
{Kobayashi}, C., {Karakas}, A.~I., \& {Lugaro}, M. 2020, \href{http://dx.doi.org/10.3847/1538-4357/abae65}{\JournalTitle{\apj}, 900, 179}

\bibitem[{{Kobayashi} {et~al.}(2011){Kobayashi}, {Karakas}, \& {Umeda}}]{Kobayashi2011b}
{Kobayashi}, C., {Karakas}, A.~I., \& {Umeda}, H. 2011, \href{http://dx.doi.org/10.1111/j.1365-2966.2011.18621.x}{\JournalTitle{\mnras}, 414, 3231}

\bibitem[{{Kobayashi} \& {Nomoto}(2009)}]{Kobayashi2009}
{Kobayashi}, C., \& {Nomoto}, K. 2009, \href{http://dx.doi.org/10.1088/0004-637X/707/2/1466}{\JournalTitle{\apj}, 707, 1466}

\bibitem[{{Kobayashi} {et~al.}(2006){Kobayashi}, {Umeda}, {Nomoto}, {Tominaga}, \& {Ohkubo}}]{Kobayashi2006}
{Kobayashi}, C., {Umeda}, H., {Nomoto}, K., {Tominaga}, N., \& {Ohkubo}, T. 2006, \href{http://dx.doi.org/10.1086/508914}{\JournalTitle{\apj}, 653, 1145}

\bibitem[{{Krumholz} \& {Ting}(2018)}]{Krumholz2018}
{Krumholz}, M.~R., \& {Ting}, Y.-S. 2018, \href{http://dx.doi.org/10.1093/mnras/stx3286}{\JournalTitle{\mnras}, 475, 2236}

\bibitem[{{Lah{\'e}n} {et~al.}(2020){Lah{\'e}n}, {Naab}, {Johansson}, {Elmegreen}, {Hu}, {Walch}, {Steinwandel}, \& {Moster}}]{Lahen+2020}
{Lah{\'e}n}, N., {Naab}, T., {Johansson}, P.~H., {et~al.} 2020, \href{http://dx.doi.org/10.3847/1538-4357/ab7190}{\JournalTitle{\apj}, 891, 2}

\bibitem[{{Lah{\'e}n} {et~al.}(2023){Lah{\'e}n}, {Naab}, {Kauffmann}, {Sz{\'e}csi}, {Hislop}, {Rantala}, {Kozyreva}, {Walch}, \& {Hu}}]{Lahen+2023}
{Lah{\'e}n}, N., {Naab}, T., {Kauffmann}, G., {et~al.} 2023, \href{http://dx.doi.org/10.1093/mnras/stad1147}{\JournalTitle{\mnras}, 522, 3092}

\bibitem[{{Lattimer} \& {Schramm}(1974)}]{Lattimer1974}
{Lattimer}, J.~M., \& {Schramm}, D.~N. 1974, \href{http://dx.doi.org/10.1086/181612}{\JournalTitle{\apjl}, 192, L145}

\bibitem[{{Li} {et~al.}(2018){Li}, {Simon}, {Pace}, {Torrealba}, {Kuehn}, {Drlica-Wagner}, {Bechtol}, {Vivas}, {van der Marel}, {Wood}, {Yanny}, {Belokurov}, {Jethwa}, {Zucker}, {Lewis}, {Kron}, {Nidever}, {S{\'a}nchez-Conde}, {Ji}, {Conn}, {James}, {Martin}, {Martinez-Delgado}, {No{\"e}l}, \& {MagLiteS Collaboration}}]{Li2018}
{Li}, T.~S., {Simon}, J.~D., {Pace}, A.~B., {et~al.} 2018, \href{http://dx.doi.org/10.3847/1538-4357/aab666}{\JournalTitle{\apj}, 857, 145}

\bibitem[{{Majewski} {et~al.}(2017){Majewski}, {Schiavon}, {Frinchaboy}, {Allende Prieto}, {Barkhouser}, {Bizyaev}, {Blank}, {Brunner}, {Burton}, {Carrera}, {Chojnowski}, {Cunha}, {Epstein}, {Fitzgerald}, {Garc{\'\i}a P{\'e}rez}, {Hearty}, {Henderson}, {Holtzman}, {Johnson}, {Lam}, {Lawler}, {Maseman}, {M{\'e}sz{\'a}ros}, {Nelson}, {Nguyen}, {Nidever}, {Pinsonneault}, {Shetrone}, {Smee}, {Smith}, {Stolberg}, {Skrutskie}, {Walker}, {Wilson}, {Zasowski}, {Anders}, {Basu}, {Beland}, {Blanton}, {Bovy}, {Brownstein}, {Carlberg}, {Chaplin}, {Chiappini}, {Eisenstein}, {Elsworth}, {Feuillet}, {Fleming}, {Galbraith-Frew}, {Garc{\'\i}a}, {Garc{\'\i}a-Hern{\'a}ndez}, {Gillespie}, {Girardi}, {Gunn}, {Hasselquist}, {Hayden}, {Hekker}, {Ivans}, {Kinemuchi}, {Klaene}, {Mahadevan}, {Mathur}, {Mosser}, {Muna}, {Munn}, {Nichol}, {O'Connell}, {Parejko}, {Robin}, {Rocha-Pinto}, {Schultheis}, {Serenelli}, {Shane}, {Silva Aguirre}, {Sobeck}, {Thompson}, {Troup}, {Weinberg}, \& {Zamora}}]{apogee_overview}
{Majewski}, S.~R., {Schiavon}, R.~P., {Frinchaboy}, P.~M., {et~al.} 2017, \href{http://dx.doi.org/10.3847/1538-3881/aa784d}{\JournalTitle{\aj}, 154, 94}

\bibitem[{{Matteucci} \& {Greggio}(1986)}]{Matteucci1986}
{Matteucci}, F., \& {Greggio}, L. 1986, \JournalTitle{\aap}, 154, 279

\bibitem[{{McConnachie}(2012)}]{McConnachie2012}
{McConnachie}, A.~W. 2012, \href{http://dx.doi.org/10.1088/0004-6256/144/1/4}{\JournalTitle{\aj}, 144, 4}

\bibitem[{{M{\'e}sz{\'a}ros} {et~al.}(2015){M{\'e}sz{\'a}ros}, {Martell}, {Shetrone}, {Lucatello}, {Troup}, {Bovy}, {Cunha}, {Garc{\'\i}a-Hern{\'a}ndez}, {Overbeek}, {Allende Prieto}, {Beers}, {Frinchaboy}, {Garc{\'\i}a P{\'e}rez}, {Hearty}, {Holtzman}, {Majewski}, {Nidever}, {Schiavon}, {Schneider}, {Sobeck}, {Smith}, {Zamora}, \& {Zasowski}}]{Meszaros+2015}
{M{\'e}sz{\'a}ros}, S., {Martell}, S.~L., {Shetrone}, M., {et~al.} 2015, \href{http://dx.doi.org/10.1088/0004-6256/149/5/153}{\JournalTitle{\aj}, 149, 153}

\bibitem[{{M{\'e}sz{\'a}ros} {et~al.}(2020){M{\'e}sz{\'a}ros}, {Masseron}, {Garc{\'\i}a-Hern{\'a}ndez}, {Allende Prieto}, {Beers}, {Bizyaev}, {Chojnowski}, {Cohen}, {Cunha}, {Dell'Agli}, {Ebelke}, {Fern{\'a}ndez-Trincado}, {Frinchaboy}, {Geisler}, {Hasselquist}, {Hearty}, {Holtzman}, {Johnson}, {Lane}, {Lacerna}, {Longa-Pe{\~n}a}, {Majewski}, {Martell}, {Minniti}, {Nataf}, {Nidever}, {Pan}, {Schiavon}, {Shetrone}, {Smith}, {Sobeck}, {Stringfellow}, {Szigeti}, {Tang}, {Wilson}, \& {Zamora}}]{Meszaros2020}
{M{\'e}sz{\'a}ros}, S., {Masseron}, T., {Garc{\'\i}a-Hern{\'a}ndez}, D.~A., {et~al.} 2020, \href{http://dx.doi.org/10.1093/mnras/stz3496}{\JournalTitle{\mnras}, 492, 1641}

\bibitem[{{Milone} \& {Marino}(2022)}]{Milone&Marino2022}
{Milone}, A.~P., \& {Marino}, A.~F. 2022, \href{http://dx.doi.org/10.3390/universe8070359}{\JournalTitle{Universe}, 8, 359}

\bibitem[{{Milone} {et~al.}(2018){Milone}, {Marino}, {Renzini}, {D'Antona}, {Anderson}, {Barbuy}, {Bedin}, {Bellini}, {Brown}, {Cassisi}, {Cordoni}, {Lagioia}, {Nardiello}, {Ortolani}, {Piotto}, {Sarajedini}, {Tailo}, {van der Marel}, \& {Vesperini}}]{Milone2018}
{Milone}, A.~P., {Marino}, A.~F., {Renzini}, A., {et~al.} 2018, \href{http://dx.doi.org/10.1093/mnras/sty2573}{\JournalTitle{\mnras}, 481, 5098}

\bibitem[{{Muley} {et~al.}(2021){Muley}, {Wheeler}, {Hopkins}, {Wetzel}, {Emerick}, \& {Kere{\v{s}}}}]{Muley2021}
{Muley}, D.~A., {Wheeler}, C.~R., {Hopkins}, P.~F., {et~al.} 2021, \href{http://dx.doi.org/10.1093/mnras/stab2572}{\JournalTitle{\mnras}, 508, 508}

\bibitem[{{Ness} {et~al.}(2018){Ness}, {Rix}, {Hogg}, {Casey}, {Holtzman}, {Fouesneau}, {Zasowski}, {Geisler}, {Shetrone}, {Minniti}, {Frinchaboy}, \& {Roman-Lopes}}]{Ness2018}
{Ness}, M., {Rix}, H.~W., {Hogg}, D.~W., {et~al.} 2018, \href{http://dx.doi.org/10.3847/1538-4357/aa9d8e}{\JournalTitle{\apj}, 853, 198}

\bibitem[{{Neumayer} {et~al.}(2020){Neumayer}, {Seth}, \& {B{\"o}ker}}]{Neumayer+2020}
{Neumayer}, N., {Seth}, A., \& {B{\"o}ker}, T. 2020, \href{http://dx.doi.org/10.1007/s00159-020-00125-0}{\JournalTitle{\aapr}, 28, 4}

\bibitem[{{Pan} \& {Kravtsov}(2023)}]{Pan2023}
{Pan}, Y., \& {Kravtsov}, A. 2023, \href{http://dx.doi.org/10.48550/arXiv.2310.08636}{\JournalTitle{arXiv e-prints}, arXiv:2310.08636}

\bibitem[{{Patel} {et~al.}(2022){Patel}, {Loebman}, {Wetzel}, {Faucher-Gigu{\`e}re}, {El-Badry}, \& {Bailin}}]{Patel2022}
{Patel}, P.~B., {Loebman}, S.~R., {Wetzel}, A., {et~al.} 2022, \href{http://dx.doi.org/10.1093/mnras/stac834}{\JournalTitle{\mnras}, 512, 5671}

\bibitem[{{Pignatari} {et~al.}(2010){Pignatari}, {Gallino}, {Heil}, {Wiescher}, {K{\"a}ppeler}, {Herwig}, \& {Bisterzo}}]{Pignatari2010}
{Pignatari}, M., {Gallino}, R., {Heil}, M., {et~al.} 2010, \href{http://dx.doi.org/10.1088/0004-637X/710/2/1557}{\JournalTitle{\apj}, 710, 1557}

\bibitem[{{Piotto} {et~al.}(2012){Piotto}, {Milone}, {Anderson}, {Bedin}, {Bellini}, {Cassisi}, {Marino}, {Aparicio}, \& {Nascimbeni}}]{Piotto2012}
{Piotto}, G., {Milone}, A.~P., {Anderson}, J., {et~al.} 2012, \href{http://dx.doi.org/10.1088/0004-637X/760/1/39}{\JournalTitle{\apj}, 760, 39}

\bibitem[{{Piotto} {et~al.}(2015){Piotto}, {Milone}, {Bedin}, {Anderson}, {King}, {Marino}, {Nardiello}, {Aparicio}, {Barbuy}, {Bellini}, {Brown}, {Cassisi}, {Cool}, {Cunial}, {Dalessandro}, {D'Antona}, {Ferraro}, {Hidalgo}, {Lanzoni}, {Monelli}, {Ortolani}, {Renzini}, {Salaris}, {Sarajedini}, {van der Marel}, {Vesperini}, \& {Zoccali}}]{Piotto2015}
{Piotto}, G., {Milone}, A.~P., {Bedin}, L.~R., {et~al.} 2015, \href{http://dx.doi.org/10.1088/0004-6256/149/3/91}{\JournalTitle{\aj}, 149, 91}

\bibitem[{{Poovelil} {et~al.}(2020){Poovelil}, {Zasowski}, {Hasselquist}, {Seth}, {Donor}, {Beaton}, {Cunha}, {Frinchaboy}, {Garc{\'\i}a-Hern{\'a}ndez}, {Hawkins}, {Kratter}, {Lane}, \& {Nitschelm}}]{Poovelil2020}
{Poovelil}, V.~J., {Zasowski}, G., {Hasselquist}, S., {et~al.} 2020, \href{http://dx.doi.org/10.3847/1538-4357/abb93e}{\JournalTitle{\apj}, 903, 55}

\bibitem[{{Rice} \& {Brewer}(2020)}]{Rice2020}
{Rice}, M., \& {Brewer}, J.~M. 2020, \href{http://dx.doi.org/10.3847/1538-4357/ab9f96}{\JournalTitle{\apj}, 898, 119}

\bibitem[{{Robertson} {et~al.}(2005){Robertson}, {Bullock}, {Font}, {Johnston}, \& {Hernquist}}]{Robertson2005}
{Robertson}, B., {Bullock}, J.~S., {Font}, A.~S., {Johnston}, K.~V., \& {Hernquist}, L. 2005, \href{http://dx.doi.org/10.1086/452619}{\JournalTitle{\apj}, 632, 872}

\bibitem[{{Rosswog} {et~al.}(1999){Rosswog}, {Liebend{\"o}rfer}, {Thielemann}, {Davies}, {Benz}, \& {Piran}}]{Rosswog1999}
{Rosswog}, S., {Liebend{\"o}rfer}, M., {Thielemann}, F.~K., {et~al.} 1999, \JournalTitle{\aap}, 341, 499

\bibitem[{{Schiavon} {et~al.}(2024){Schiavon}, {Phillips}, {Myers}, {Horta}, {Minniti}, {Allende Prieto}, {Anguiano}, {Beaton}, {Beers}, {Brownstein}, {Cohen}, {Fern{\'a}ndez-Trincado}, {Frinchaboy}, {J{\"o}nsson}, {Kisku}, {Lane}, {Majewski}, {Mason}, {M{\'e}sz{\'a}ros}, \& {Stringfellow}}]{GC_VAC_APOGEE_DR17}
{Schiavon}, R.~P., {Phillips}, S.~G., {Myers}, N., {et~al.} 2024, \href{http://dx.doi.org/10.1093/mnras/stad3020}{\JournalTitle{\mnras}, 528, 1393}

\bibitem[{{Smith} {et~al.}(2021){Smith}, {Bizyaev}, {Cunha}, {Shetrone}, {Souto}, {Allende Prieto}, {Masseron}, {M{\'e}sz{\'a}ros}, {J{\"o}nsson}, {Hasselquist}, {Osorio}, {Garc{\'\i}a-Hern{\'a}ndez}, {Plez}, {Beaton}, {Holtzman}, {Majewski}, {Stringfellow}, \& {Sobeck}}]{apogee_specline}
{Smith}, V.~V., {Bizyaev}, D., {Cunha}, K., {et~al.} 2021, \href{http://dx.doi.org/10.3847/1538-3881/abefdc}{\JournalTitle{\aj}, 161, 254}

\bibitem[{{Strassmeier} {et~al.}(2015){Strassmeier}, {Ilyin}, {J{\"a}rvinen}, {Weber}, {Woche}, {Barnes}, {Bauer}, {Beckert}, {Bittner}, {Bredthauer}, {Carroll}, {Denker}, {Dionies}, {DiVarano}, {D{\"o}scher}, {Fechner}, {Feuerstein}, {Granzer}, {Hahn}, {Harnisch}, {Hofmann}, {Lesser}, {Paschke}, {Pankratow}, {Plank}, {Pl{\"u}schke}, {Popow}, \& {Sablowski}}]{PEPSI2015}
{Strassmeier}, K.~G., {Ilyin}, I., {J{\"a}rvinen}, A., {et~al.} 2015, \href{http://dx.doi.org/10.1002/asna.201512172}{\JournalTitle{Astronomische Nachrichten}, 336, 324}

\bibitem[{{Timmes} {et~al.}(1995){Timmes}, {Woosley}, \& {Weaver}}]{Timmes1995}
{Timmes}, F.~X., {Woosley}, S.~E., \& {Weaver}, T.~A. 1995, \href{http://dx.doi.org/10.1086/192172}{\JournalTitle{\apjs}, 98, 617}

\bibitem[{{Tinsley}(1980)}]{Tinsley1980}
{Tinsley}, B.~M. 1980, \href{http://dx.doi.org/10.48550/arXiv.2203.02041}{\JournalTitle{\fcp}, 5, 287}

\bibitem[{{Unavane} {et~al.}(1996){Unavane}, {Wyse}, \& {Gilmore}}]{Unavane1996}
{Unavane}, M., {Wyse}, R. F.~G., \& {Gilmore}, G. 1996, \href{http://dx.doi.org/10.1093/mnras/278.3.727}{\JournalTitle{\mnras}, 278, 727}

\bibitem[{{Venn} {et~al.}(2004){Venn}, {Irwin}, {Shetrone}, {Tout}, {Hill}, \& {Tolstoy}}]{Venn2004}
{Venn}, K.~A., {Irwin}, M., {Shetrone}, M.~D., {et~al.} 2004, \href{http://dx.doi.org/10.1086/422734}{\JournalTitle{\aj}, 128, 1177}

\bibitem[{{Walker} {et~al.}(2006){Walker}, {Mateo}, {Olszewski}, {Bernstein}, {Wang}, \& {Woodroofe}}]{Walker2006}
{Walker}, M.~G., {Mateo}, M., {Olszewski}, E.~W., {et~al.} 2006, \href{http://dx.doi.org/10.1086/500193}{\JournalTitle{\aj}, 131, 2114}

\bibitem[{{Wanajo} {et~al.}(2014){Wanajo}, {Sekiguchi}, {Nishimura}, {Kiuchi}, {Kyutoku}, \& {Shibata}}]{Wanajo2014}
{Wanajo}, S., {Sekiguchi}, Y., {Nishimura}, N., {et~al.} 2014, \href{http://dx.doi.org/10.1088/2041-8205/789/2/L39}{\JournalTitle{\apjl}, 789, L39}

\bibitem[{{Weinberg} {et~al.}(2022){Weinberg}, {Holtzman}, {Johnson}, {Hayes}, {Hasselquist}, {Shetrone}, {Ting}, {Beaton}, {Beers}, {Bird}, {Bizyaev}, {Blanton}, {Cunha}, {Fern{\'a}ndez-Trincado}, {Frinchaboy}, {Garc{\'\i}a-Hern{\'a}ndez}, {Griffith}, {Johnson}, {J{\"o}nsson}, {Lane}, {Leung}, {Mackereth}, {Majewski}, {M{\'e}sz{\'a}ros}, {Nitschelm}, {Pan}, {Schiavon}, {Schneider}, {Schultheis}, {Smith}, {Sobeck}, {Stassun}, {Stringfellow}, {Vincenzo}, {Wilson}, \& {Zasowski}}]{Weinberg2022}
{Weinberg}, D.~H., {Holtzman}, J.~A., {Johnson}, J.~A., {et~al.} 2022, \href{http://dx.doi.org/10.3847/1538-4365/ac6028}{\JournalTitle{\apjs}, 260, 32}

\bibitem[{{Welsh} {et~al.}(2021){Welsh}, {Cooke}, \& {Fumagalli}}]{Welsh2021}
{Welsh}, L., {Cooke}, R., \& {Fumagalli}, M. 2021, \href{http://dx.doi.org/10.1093/mnras/staa3342}{\JournalTitle{\mnras}, 500, 5214}

\bibitem[{{Xiang} {et~al.}(2021){Xiang}, {Rix}, {Ting}, {Zari}, {El-Badry}, {Yuan}, \& {Cui}}]{Xiang2021}
{Xiang}, M., {Rix}, H.-W., {Ting}, Y.-S., {et~al.} 2021, \href{http://dx.doi.org/10.3847/1538-4365/abd6ba}{\JournalTitle{\apjs}, 253, 22}

\bibitem[{{Xiang} {et~al.}(2022){Xiang}, {Rix}, {Ting}, {Kudritzki}, {Conroy}, {Zari}, {Shi}, {Przybilla}, {Ramirez-Tannus}, {Tkachenko}, {Gebruers}, \& {Liu}}]{Xiang2022}
---. 2022, \href{http://dx.doi.org/10.1051/0004-6361/202141570}{\JournalTitle{\aap}, 662, A66}

\bibitem[{{Yong} {et~al.}(2013){Yong}, {Mel{\'e}ndez}, {Grundahl}, {Roederer}, {Norris}, {Milone}, {Marino}, {Coelho}, {McArthur}, {Lind}, {Collet}, \& {Asplund}}]{Yong2013}
{Yong}, D., {Mel{\'e}ndez}, J., {Grundahl}, F., {et~al.} 2013, \href{http://dx.doi.org/10.1093/mnras/stt1276}{\JournalTitle{\mnras}, 434, 3542}

\end{thebibliography}

\appendix
\label{sec:appendix}
\begin{deluxetable}{c||ccccccccccccc}[h!]
\tabletypesize{\scriptsize}
\tablecaption{(1) Object Name; (2) Total number of stars that meet initial data cuts; (3) V-band Magnitude\footnote{From \citet{McConnachie2012} for dSphs and \citet{Harris1996} (2010 edition) for GCs}; (4) Average [Fe/H] of stars in cut; (5) Median \textbf{S/N} of stars in cut; (6-12) Intrinsic scatter $\pm$ standard deviation \textbf{(dex)} of [X/Fe] for stars in cut.}
\tablenum{3} \label{tab:appendix}
\tablehead{
\colhead{Object} & \colhead{ Stars }& \colhead{$M_V$} & \colhead{$\overline{[Fe/H]}$} & \colhead{S/N} & \colhead{$\sigma_{[Al/Fe]}$} & \colhead{$\sigma_{[O/Fe]}$} & \colhead{$\sigma_{[Mg/Fe]}$} & \colhead{$\sigma_{[Si/Fe]}$} & \colhead{$\sigma_{[Ti/Fe]}$} & \colhead{$\sigma_{[Ni/Fe]}$} & \colhead{$\sigma_{[Mn/Fe]}$} \\
\hline
\textit{halo}& & & & & & \\
\textbf{MW Halo}    & 291   & N/A   & -1.31 & 260.1 &   $0.10 \pm 0.01$ & $0.10 \pm 0.01$ & $0.05 \pm 0.00$ & $0.07 \pm 0.01$ & $0.04 \pm 0.02$ & $0.04 \pm 0.00$ & $0.06 \pm 0.01$ \\
\hline
\textit{dSph}& & & & & & \\
\textbf{Draco}      & 31    & -8.8  & -1.74 & 78.4  &   $0.06 \pm 0.05$ & $0.06 \pm 0.05$ & $0.15 \pm 0.05$ & $0.06 \pm 0.04$ & $0.05 \pm 0.04$ & $0.07 \pm 0.05$ & $0.06 \pm 0.05$ \\
\textbf{Ursa Minor} & 40    & -8.8  & -1.84 & 48.9  &   $0.05 \pm 0.04$ & $0.05 \pm 0.04$ & $0.05 \pm 0.04$ & $0.05 \pm 0.04$ & $0.06 \pm 0.05$ & $0.15 \pm 0.05$ & $0.14 \pm 0.08$ \\
\textbf{Sextans}    & 27    & -9.3  & -1.74 & 54.3  &   $0.06 \pm 0.05$ & $0.06 \pm 0.05$ & $0.07 \pm 0.05$ & $0.06 \pm 0.05$ & $0.14 \pm 0.09$ & $0.07 \pm 0.05$ & $0.10 \pm 0.08$ \\
\textbf{Fornax}     & 62    & -13.4 & -1.28 & 57.1  &   $0.04 \pm 0.03$ & $0.03 \pm 0.02$ & $0.03 \pm 0.02$ & $0.03 \pm 0.02$ & $0.04 \pm 0.03$ & $0.04 \pm 0.03$ & $0.03 \pm 0.03$ \\
\textbf{Sculptor}   & 126   & -11.1 & -1.57 & 57.5  &   $0.03 \pm 0.02$ & $0.02 \pm 0.02$ & $0.03 \pm 0.02$ & $0.02 \pm 0.02$ & $0.06 \pm 0.03$ & $0.13 \pm 0.02$ & $0.04 \pm 0.03$ \\
\textbf{Carina}     & 47    & -9.1  & -1.59 & 68.6  &   $0.05 \pm 0.04$ & $0.04 \pm 0.03$ & $0.05 \pm 0.04$ & $0.04 \pm 0.03$ & $0.04 \pm 0.03$ & $0.04 \pm 0.03$ & $0.05 \pm 0.04$ \\
\hline
\textit{Undetermined}& & & & & & \\
\textbf{OmegaCen}   & 1211  & -10.26& -1.63 & 153.0 &   $0.10 \pm 0.01$ & $0.17 \pm 0.01$ & $0.11 \pm 0.00$ & $0.03 \pm 0.00$ & $0.11 \pm 0.01$ & $0.03 \pm 0.00$ & $0.14 \pm 0.01$ \\
\textbf{M54}        & 40    & -9.98 & -1.46 & 88.6  &   $0.04 \pm 0.03$ & $0.05 \pm 0.03$ & $0.02 \pm 0.01$ & $0.02 \pm 0.02$ & $0.14 \pm 0.15$ & $0.02 \pm 0.01$ & $0.04 \pm 0.03$ \\
\hline
\textit{GC}& & & & & & \\
\textbf{M53}        & 34    & -8.71 & -1.89 & 131.9 &   $0.09 \pm 0.07$ & $0.03 \pm 0.03$ & $0.04 \pm 0.02$ & $0.04 \pm 0.03$ & $0.13 \pm 0.06$ & $0.05 \pm 0.04$ & $0.10 \pm 0.06$ \\
\textbf{N5466}      & 14    & -6.98 & -1.81 & 71.2  &   $0.15 \pm 0.10$ & $0.06 \pm 0.06$ & $0.05 \pm 0.05$ & $0.08 \pm 0.06$ & $0.11 \pm 0.10$ & $0.14 \pm 0.06$ & $0.10 \pm 0.08$ \\
\textbf{M2}         & 28    & -9.03 & -1.48 & 137.7 &   $0.15 \pm 0.05$ & $0.03 \pm 0.02$ & $0.05 \pm 0.01$ & $0.02 \pm 0.02$ & $0.04 \pm 0.03$ & $0.02 \pm 0.01$ & $0.03 \pm 0.02$ \\
\textbf{M13}        & 90    & -8.55 & -1.48 & 134.4 &   $0.26 \pm 0.03$ & $0.10 \pm 0.02$ & $0.08 \pm 0.01$ & $0.01 \pm 0.01$ & $0.02 \pm 0.01$ & $0.06 \pm 0.01$ & $0.15 \pm 0.02$ \\
\textbf{M3}         & 236   & -8.88 & -1.42 & 148.0 &   $0.11 \pm 0.01$ & $0.06 \pm 0.01$ & $0.05 \pm 0.00$ & $0.01 \pm 0.01$ & $0.24 \pm 0.02$ & $0.04 \pm 0.01$ & $0.08 \pm 0.01$ \\
\textbf{M5}         & 147   & -8.81 & -1.22 & 166.3 &   $0.17 \pm 0.02$ & $0.14 \pm 0.01$ & $0.05 \pm 0.00$ & $0.01 \pm 0.01$ & $0.01 \pm 0.01$ & $0.02 \pm 0.01$ & $0.01 \pm 0.01$ \\
\textbf{M12}        & 75    & -7.31 & -1.28 & 125.5 &   $0.05 \pm 0.01$ & $0.03 \pm 0.01$ & $0.02 \pm 0.01$ & $0.01 \pm 0.01$ & $0.02 \pm 0.02$ & $0.01 \pm 0.01$ & $0.11 \pm 0.03$ \\
\textbf{N6397}      & 56    & -6.64 & -2.00 & 360.4 &   $0.22 \pm 0.05$ & $0.03 \pm 0.02$ & $0.04 \pm 0.01$ & $0.01 \pm 0.01$ & $0.11 \pm 0.04$ & $0.04 \pm 0.02$ & $0.15 \pm 0.04$ \\
\textbf{M55}        & 83    & -7.57 & -1.77 & 181.3 &   $0.21 \pm 0.04$ & $0.09 \pm 0.02$ & $0.02 \pm 0.01$ & $0.01 \pm 0.01$ & $0.03 \pm 0.02$ & $0.02 \pm 0.02$ & $0.08 \pm 0.03$ \\
\textbf{M22}        & 231   & -8.50 & -1.69 & 140.1 &   $0.23 \pm 0.03$ & $0.09 \pm 0.01$ & $0.06 \pm 0.01 $& $0.01 \pm 0.01$ & $0.20 \pm 0.02$ & $0.08 \pm 0.01$ & $0.22 \pm 0.02$ \\
\textbf{M79}        & 31    & -7.86 & -1.51 & 166.3 &   $0.07 \pm 0.07$ & $0.02 \pm 0.02$ & $0.05 \pm 0.01$ & $0.02 \pm 0.01$ & $0.04 \pm 0.03$ & $0.17 \pm 0.04$ & $0.13 \pm 0.04$ \\
\textbf{N3201}      & 101   & -7.45 & -1.35 & 192.9 &   $0.07 \pm 0.01$ & $0.14 \pm 0.01$ & $0.03 \pm 0.01$ & $0.01 \pm 0.00$ & $0.02 \pm 0.01$ & $0.02 \pm 0.01$ & $0.04 \pm 0.02$ \\
\textbf{M10}        & 83    & -7.48 & -1.51 & 176.3 &   $0.04 \pm 0.02$ & $0.11 \pm 0.02$ & $0.08 \pm 0.01$ & $0.01 \pm 0.01$ & $0.03 \pm 0.02$ & $0.02 \pm 0.01$ & $0.05 \pm 0.03$ \\
\textbf{N6752}      & 79    & -7.73 & -1.49 & 202.0 &   $0.16 \pm 0.03$ & $0.15 \pm 0.02$ & $0.06 \pm 0.01$ & $0.01 \pm 0.01$ & $0.02 \pm 0.02$ & $0.01 \pm 0.01$ & $0.11 \pm 0.02$ \\
\textbf{N288}       & 35    & -6.75 & -1.27 & 150.0 &   $0.03 \pm 0.03$ & $0.02 \pm 0.01$ & $0.02 \pm 0.01$ & $0.02 \pm 0.01$ & $0.05 \pm 0.04$ & $0.01 \pm 0.01$ & $0.03 \pm 0.02$ \\
\textbf{N362}       & 58    & -8.43 & -1.11 & 130.9 &   $0.07 \pm 0.02$ & $0.04 \pm 0.01$ & $0.04 \pm 0.01$ & $0.01 \pm 0.01$ & $0.02 \pm 0.02$ & $0.03 \pm 0.01$ & $0.02 \pm 0.01$ \\
\textbf{N1851}      & 34    & -8.33 & -1.11 & 128.3 &   $0.12 \pm 0.03$ & $0.02 \pm 0.01$ & $0.01 \pm 0.01$ & $0.02 \pm 0.01$ & $0.04 \pm 0.03$ & $0.01 \pm 0.01$ & $0.02 \pm 0.02$ \\
\textbf{M4}         & 109   & -7.19 & -1.05 & 222.4 &   $0.03 \pm 0.02$ & $0.02 \pm 0.01$ & $0.02 \pm 0.00$ & $0.01 \pm 0.00$ & $0.02 \pm 0.02$ & $0.00 \pm 0.00$ & $0.01 \pm 0.01$ \\
\textbf{N2808}      & 80    & -9.39 & -1.11 & 148.2 &   $0.05 \pm 0.01$ & $0.08 \pm 0.01$ & $0.04 \pm 0.01$ & $0.01 \pm 0.01$ & $0.02 \pm 0.02$ & $0.01 \pm 0.00$ & $0.01 \pm 0.01$
}
\startdata
\enddata

\end{deluxetable}

\end{document}